\documentclass[a4paper]{article}

\usepackage[english]{babel}
\usepackage[utf8x]{inputenc}
\usepackage[T1]{fontenc}

\usepackage[a4paper,top=3cm,bottom=2cm,left=2.5cm,right=2.5cm,marginparwidth=1.75cm]{geometry}

\usepackage{fancyhdr}
\usepackage{amssymb}
\usepackage{amsmath}
\usepackage{graphicx}
\usepackage[colorinlistoftodos]{todonotes}
\usepackage[colorlinks=true, allcolors=blue]{hyperref}

\usepackage[authoryear,square]{natbib}
\bibpunct{(}{)}{;}{a}{}{,}
\usepackage{etoolbox}
\usepackage{subcaption}
\usepackage{caption}
\usepackage{wrapfig, placeins}
\usepackage{authblk}
\usepackage{ulem}

\title{MeerKLASS: MeerKAT Large Area Synoptic Survey}

\newcommand{\skipthis}[1]{}

\newcommand{\apjl}{ApJ}
\newcommand{\apjs}{ApJS}
\newcommand{\aap}{A \& A}

\newcommand{\be}{\begin{equation}}
\newcommand{\ee}{\end{equation}}
\newcommand{\bea}{\begin{eqnarray}}
\newcommand{\eea}{\end{eqnarray}}

\DeclareRobustCommand{\ion}[2]{\textup{#1\,\textsc{\lowercase{#2}}}}

\author[1,2]{Mario G. Santos\thanks{Email: mgrsantos@uwc.ac.za}}
\author[1,3]{Michelle Cluver} 
\author[4]{Matt Hilton} 
\author[5,1]{Matt Jarvis} 
\author[2,6,7]{Gyula I. G. Jozsa}
\author[8]{Lerothodi Leeuw} 
\author[2,6]{Oleg Smirnov} 
\author[1,3,9]{Russ Taylor}
\author[10,6]{Filipe Abdalla}
\author[11,12]{Jose Afonso}
\author[5]{David Alonso}
\author[13]{David Bacon}
\author[14,15,16]{Bruce A. Bassett}
\author[2,6]{Gianni Bernardi}
\author[17,18]{Philip Bull}
\author[19,20,21,1]{Stefano Camera}
\author[4]{H. Cynthia Chiang}
\author[22]{Sergio Colafrancesco}
\author[5]{Pedro G. Ferreira}
\author[1]{Jose Fonseca}
\author[9]{Kurt van der Heyden}
\author[5,6]{Ian Heywood}	
\author[23]{Kenda Knowles}
\author[2,14]{Michelle Lochner}
\author[23,24,25]{Yin-Zhe Ma}
\author[1,13]{Roy Maartens}
\author[2,6]{Sphesihle Makhathini}
\author[4]{Kavilan Moodley}
\author[13,26]{Alkistis Pourtsidou}
\author[1]{Matthew Prescott}
\author[23]{Jonathan Sievers}
\author[27]{Kristine Spekkens}
\author[1,28]{Mattia Vaccari}
\author[16]{Amanda Weltman}
\author[1]{Imogen Whittam}
\author[1,16]{Amadeus Witzemann}
\author[29]{Laura Wolz} 
\author[1,9]{Jonathan T. L. Zwart}
\affil[1]{Department of Physics and Astronomy, University of the Western Cape, Cape Town 7535, South Africa}
\affil[2]{SKA South Africa, 3rd Floor, The Park, Park Road, Pinelands, 7405, South Africa}
\affil[3]{Inter-University Institute for Data Intensive Astronomy, University of Cape Town, Rondebosch, 7701, South Africa}
\affil[4]{School of Mathematics, Statistics, and Computer Science, Astrophysics and Cosmology Research Unit, University of KwaZulu-Natal, Durban, 4041, South Africa}
\affil[5]{Astrophysics, University of Oxford, Keble Road, Oxford OX1 3RH UK}
\affil[6]{Department of Physics and Electronics, Rhodes University, PO BOX 94, Grahamstown, 6140, South Africa}
\affil[7]{Argelander-Institut f\"ur Astronomie, Auf dem H\"ugel 71, D-53121 Bonn, Germany}
\affil[8]{UNISA, South Africa}
\affil[9]{Department of Astronomy, University of Cape Town, Cape Town, 7005 South Africa}
\affil[10]{Department of Physics \& Astronomy, University College London, Gower Street, London, WC1E 6BT, UK}
\affil[11]{Instituto de Astrof\'{i}sica e Ci\^{e}ncias do Espa\c co, Universidade de Lisboa, OAL, Tapada da Ajuda, PT1349-018 Lisboa, Portugal}
\affil[12]{Departamento de F\'{i}sica, Faculdade de Ci\^{e}ncias, Universidade de Lisboa, Edif\'{i}cio C8, Campo Grande, PT1749-016 Lisbon, Portugal}
\affil[13]{Institute of Cosmology \& Gravitation, University of Portsmouth, Burnaby Road, Portsmouth PO1 3FX, UK}
\affil[14]{African Institute for Mathematical Sciences, Muizenberg, Cape Town, South Africa} 
\affil[15]{South African Astronomical Observatory, Observatory, Cape Town, South Africa}
\affil[16]{Department of Mathematics and Applied Mathematics, University of Cape Town, Cape Town, South Africa}
\affil[17]{California Institute of Technology, Pasadena, CA 91125, USA}
\affil[18]{Jet Propulsion Laboratory, California Institute of Technology, 4800 Oak Grove Drive, Pasadena, California, USA}
\affil[19]{Dipartimento di Fisica, Universit\`a degli Studi di Torino, 10125 Torino, Italy}
\affil[20]{INFN -- Istituto Nazionale di Fisica Nucleare, Sezione di Torino, 10125 Torino, Italy}
\affil[21]{INAF -- Istituto Nazionale di Astrofisica, Osservatorio Astrofisico di Torino, 10025 Pino Torinese, Italy}
\affil[22]{School of Physics, Wits University, Johannesburg WITS-2001, South Africa}
\affil[23]{School of Chemistry and Physics, Astrophysics and Cosmology Research Unit, University of KwaZulu-Natal, Durban, 4041, South Africa}
\affil[24]{NAOC-UKZN Computational Astrophysics Centre (NUCAC), University of KwaZulu-Natal, Durban, 4000 South Africa}
\affil[25]{Purple Mountain Observatory, Chinese Academy of Sciences, Nanjing 210008, China}
\affil[26]{School of Physics and Astronomy, Queen Mary University of London, Mile End Road, London E1 4NS, UK}
\affil[27]{Royal Military College of Canada, PO Box 17000, Station Forces, Kingston, Ontario K7L 2E1, Canada}
\affil[28]{INAF - Istituto di Radioastronomia, via Gobetti 101, 40129 Bologna, Italy}
\affil[29]{School of Physics, University of Melbourne, Parkville, VIC 3010, Australia}

\date{}

\begin{document}

\maketitle

\newpage

\vspace*{3cm}

\abstract{
We discuss the ground-breaking science that will be possible with a wide area survey, using the MeerKAT telescope, known as MeerKLASS (MeerKAT Large Area Synoptic Survey). The current specifications of MeerKAT make it a great fit for science applications that require large survey speeds but not necessarily high angular resolutions. In particular, for cosmology, a large survey over $\sim 4,000 \, {\rm deg}^2$ for $\sim 4,000$ hours will potentially provide the first ever measurements of the baryon acoustic oscillations using the 21cm intensity mapping technique, with enough accuracy to impose constraints on the nature of dark energy. The combination with multi-wavelength data will give unique additional information, such as exquisite constraints on primordial non-Gaussianity using the multi-tracer technique, as well as a better handle on foregrounds and systematics. Such a wide survey with MeerKAT is also a great match for \ion{H}{i} galaxy studies. In fact, MeerKLASS can provide unrivalled statistics in the pre-SKA era, for galaxies resolved in the \ion{H}{i} emission line beyond local structures at z > 0.01. It uniquely traces the faint neutral gas with high signal to noise, in a significant statistical sample of resolved galaxies in the \ion{H}{i} mass regime of $\sim 10^9\,M_\odot$, well beyond the knee of the \ion{H}{i} mass function, providing statistics of the most abundant galaxies. It will also produce a large continuum galaxy sample down to a depth of about 5\,$\mu$Jy in L-band, which is quite unique over such large areas and will allow studies of the large-scale structure of the Universe out to high redshifts, complementing the galaxy \ion{H}{i} survey to form a transformational multi-wavelength approach to study galaxy dynamics and evolution. Finally, the same survey will supply unique information for a range of other science applications, including a large statistical investigation of galaxy clusters, and the discovery of rare high-redshift AGN that can be used to probe the epoch of reionization as well as produce a rotation measure map across a huge swathe of the sky. The MeerKLASS survey will be a crucial step on the road to using SKA1-MID for cosmological applications and other commensal surveys, as described in the top priority SKA key science projects.}

\newpage

\vspace*{2cm}

\tableofcontents

\newpage

\section{Rationale}

A new generation of radio telescopes with unprecedented capabilities for astronomy and tests of fundamental physics will enter operation over the next few years. Thanks to their large survey speeds, they present a unique opportunity to probe the Universe using large sky surveys. These will deliver measurements of the large-scale structure of the Universe across cosmic time, a crucial ingredient in pushing the current limits of precision cosmology and exploring new cosmological models, thanks to their huge 3-dimensional information content.

The key example is SKA1-MID, the first phase of the SKA, to be set in South Africa, which will deliver competitive and transformational cosmology through a series of measurements using a large sky survey \citep{Maartens:2015mra,2015aska.confE..19S, Jarvis:2015asa,Abdalla:2015zra}. However, in the immediate future, MeerKAT\footnote{https://www.ska.ac.za/science-engineering/meerkat}, the SKA precursor in South Africa, will have the capability to produce high impact cosmological constraints from a large sky survey. Several factors contribute to this: {\bf i)} the final MeerKAT specifications deliver excellent survey speed (low noise, large primary beam), exceeding initial expectations; {\bf ii)} shorter baselines make MeerKAT a better fit for large, shallow surveys, instead of deep surveys that may suffer from confusion noise; {\bf iii)} MeerKAT's dish design, with single pixel feeds, provides a stable, well-tested set-up that should allow for a better handle on calibration issues, an important consideration for extracting the small cosmological signal (in particular using the auto-correlations). A large amount of the science that we plan to do will benefit greatly from a multi-wavelength approach. Therefore, ``full sky'' surveys are not a strict requirement, as we can achieve most of the science with smaller areas that are more easily covered simultaneously by surveys at other wavelengths. 

Finally, the proposed survey  will pave the way for using the full SKA1-MID array for cosmology -- one of the SKA's key science projects. This is also true for the wide range of commensal science we are proposing here, which promise unique transformational results. The large requests of time planned with SKA1 need to be efficiently used in a commensal way by providing game changing results in different fronts, something that we will be testing with the proposed survey.

\section{Executive Summary}

In recognition of MeerKAT's potential as a cosmological survey instrument, we propose a survey of $\sim 4,000\,{\rm deg}^2$ in $\sim 4,000\,$hr, probably over the same sky area as the Dark Energy Survey, DES (\href{link}{www.darkenergysurvey.org}), which will provide a wealth of multi-wavelength data (note however that the final survey footprint is still being decided and we have also considered other possibilities such as the VIKING/KIDS area).  This multi-wavelength coverage is not only beneficial for MeerKLASS but also for other surveys that overlap in the same sky area. For instance, cross-correlations of the HI Intensity mapping (IM) survey with DES galaxies can significantly improve the photometric redshift uncertainties of these galaxies \citep{2017PhRvD..96d3515A}, something that can be crucial in improving the weak lensing constraints.

MeerKLASS will allow for a variety of scientific breakthroughs in a completely new regime: \\ 
{\bf 1.} Possibly the first ever detection of baryon acoustic oscillations (BAO) and redshift space distortions (RSD) through the 21cm intensity mapping technique, which will allow novel cosmological constraints to be placed on dark energy and alternative theories of gravity. This will be achieved in both auto-correlation and cross-correlation with other surveys, the latter guaranteeing a much cleaner signal with regard to systematics. Besides cosmology, these measurements will also constrain the HI mass function and its evolution, as well as the clustering bias of HI galaxies. \\ 
{\bf 2.} Novel constraints on primordial non-Gaussianity using the multi-tracer technique, improving on current limits. \\ 
{\bf 3.} Detect $\sim 10^7$ (mostly star-forming) galaxies in continuum, that can be used for several statistical tests in cosmology and galaxy evolution.\\ 
{\bf 4.} Detect in excess of $\sim 10^5$ galaxies directly in \ion{H}{i} providing the best statistics of resolved \ion{H}{i} galaxy observations among all planned SKA progenitor surveys, beyond the local structures at redshift z > 0.01, uniquely including a significant statistical sample of resolved galaxies within the $M_{\rm \ion{H}{i}}\,\sim\,10^9\,M_\odot$ regime at a sensitivity capable of tracing faint, extended components.\\
{\bf 5.} Probe the AGN/ SF galaxy populations up to large masses, and even find the rarest high-redshift AGN, which can be used to probe the epoch of reonisation through 21cm absorption. \\ 
{\bf 6.} Provide a large  sample of galaxy clusters and a study of radio halos and relics in clusters. \\ 
{\bf 7.} Detect thousands of sources which can be used to investigate associated HI absorption systems. \\
{\bf 8.} Produce a rotation measure map across a large patch of the sky. \\ 
{\bf 9.} Detect or set strong upper limits to the dark matter induced radio emission in nearby structures: dwarf galaxies and galaxy clusters with offset DM-baryon spatial distribution.\\
{\bf 10.} Perform slow transient searches.

MeerKLASS plans to use MeerKAT's L-band receivers (900--1670\,MHz), in order to achieve a point source sensitivity in continuum of $\sim 5.3\,\mu$Jy. 
This is assuming about 25\% band loss to RFI, and taking into account the expected confusion (which should be low at this level).
The specifications used in our calculations can be found at: \href{link}{http://meerkat2016.ska.ac.za/Files}, where we used the full 64 dishes. In the following analysis, we will sometimes consider other options (such as UHF-band or different survey areas/time) in order to justify our choices.

We aim to use the single-dish auto-correlations and interferometric data from MeerKAT simultaneously. The single-dish survey will require a particular calibration approach. The current idea is to take a fast scanning strategy, in order to suppress the effect of gain variations on the required angular scales for cosmology. The speed of such survey depends on current stability tests being done (ultimately could be just a drift scan), but we expect to take the scans at constant elevation in order to deal with ``ground pick-up''.
Second, we plan to use the MeerKAT noise diodes to inject a frequent, stable signal that can be used for calibration. The interferometer data will be taken at the same time, using an on-the-fly observing mode. This will require a careful calibration technique, where for instance, the full dish primary beam will be included. Note also that cross-correlating the intensity between dishes can help with systematics and the noise bias.
In the next sections, we describe in detail the science that can be done, and the reasoning for the requested sky area and sensitivity.

\section{Cosmology}

MeerKLASS aims to provide some of the first constraints on cosmology using the new generation of radio telescopes. With the same data, we will effectively run two different cosmological surveys: HI intensity mapping and a continuum radio galaxy survey. Combined with multi-wavelength information, these will allow exquisite constraints on several key cosmological parameters. 

\subsection{HI intensity mapping}

Typically, large-scale structure probes in cosmology use galaxy surveys with spectroscopic or photometric redshifts in the optical or near-infrared. These are threshold surveys in that they set a minimum flux above which galaxies can be individually detected. For the radio, we will have to rely on the HI 21cm signal to obtain such redshifts. But this emission line is quite weak. The next generation of radio telescopes will only achieve the required sensitivities to detect HI galaxies over small volumes not useful for cosmology. In fact, we will have to wait until SKA2 to achieve the  sensitivity to do game-changing cosmology with an HI galaxy survey \citep{abdalla}.

The intensity mapping technique relies on observations of the sky intensity over a large number of pixels instead of trying to detect individual galaxies. For a reasonably large 3D pixel (in solid angle and frequency interval), we expect to have several HI galaxies in each pixel so that their combined emission will provide a larger signal. Moreover we can use statistical techniques, similar to those that have been applied to CMB experiments, to measure quantities in the low signal to noise regime. By not requiring the detection of individual galaxies, the sensitivity requirements are much less demanding. Intensity mapping transfers the problem to one of foreground cleaning: how to develop cleaning methods to remove everything that is not the HI signal at a given frequency, which 
also impacts on the calibration requirements of the instrument (see \citealt{2015aska.confE..19S}).


Intensity mapping experiments can be classified into two types: single-dish surveys and interferometers. In single-dish surveys (e.g. using auto-correlations), each pointing of the telescope gives us one single pixel on the sky (though more dishes or feeds can be used to increase the field of view). This has the advantage of giving us the large-scale modes by scanning the sky. Since brightness temperature is independent of dish size, we can achieve the same sensitivity with a smaller dish, although that will in turn limit the angular resolution of the experiment (e.g., a 30 arcmin resolution at $z\sim 1$ would require a dish of about 50\,m in diameter). One example is the Green Bank Telescope (GBT) which produced the first (tentative) detection of the cosmological HI intensity signal by cross-correlating with the WiggleZ redshift survey \citep{Chang:2010jp,2013MNRAS.434L..46S,2013ApJ...763L..20M}. BINGO \citep{Battye:2012tg} is a proposed 40\,m multi-receiver single-dish telescope to be situated in South America and aimed at detecting the HI signal at $z\sim 0.3$.

Interferometers basically measure the Fourier transform modes of the sky. They have the advantage of easily providing high angular resolution as well being less sensitive to systematics that can plague the auto-correlation power. On the other hand, the minimum angular scale they can probe is set by their shortest baseline, which can be a problem when probing BAO scales. Some examples of purpose-built interferometer for intensity mapping are CHIME \citep{2014SPIE.9145E..22B}, aimed at detecting the BAO at $z\sim 1$, TIANLAI \citep{2015ApJ...798...40X} , set in China and HIRAX, to be built in South Africa \citep{2016SPIE.9906E..5XN}.

The next generation of large dish arrays can also potentially be exploited for HI intensity mapping measurements. This is the case for MeerKAT and SKA1-MID. However, these interferometers do not provide enough baselines on the scales of interest (5 to 80\,m), so that their sensitivity to the BAO will be small. The alternative is to use instead the auto-correlation information from each dish, e.g. with a survey using the array in single-dish mode. The large number of dishes available with these telescopes will guarantee a high survey speed for probing the HI signal. It will also allow us to probe very large cosmological scales, which will be unique to this approach. Our proposal is to use the calibrated single dish auto-correlation from MeerKAT to measure the HI IM signal, while at the same time keeping the interferometer data for the commensal science. In the following sections, we discuss what can be achieved with such survey.



\subsubsection{Detecting the HI intensity mapping signal with MeerKAT}

\begin{figure}
\centering
\includegraphics[scale=0.5]{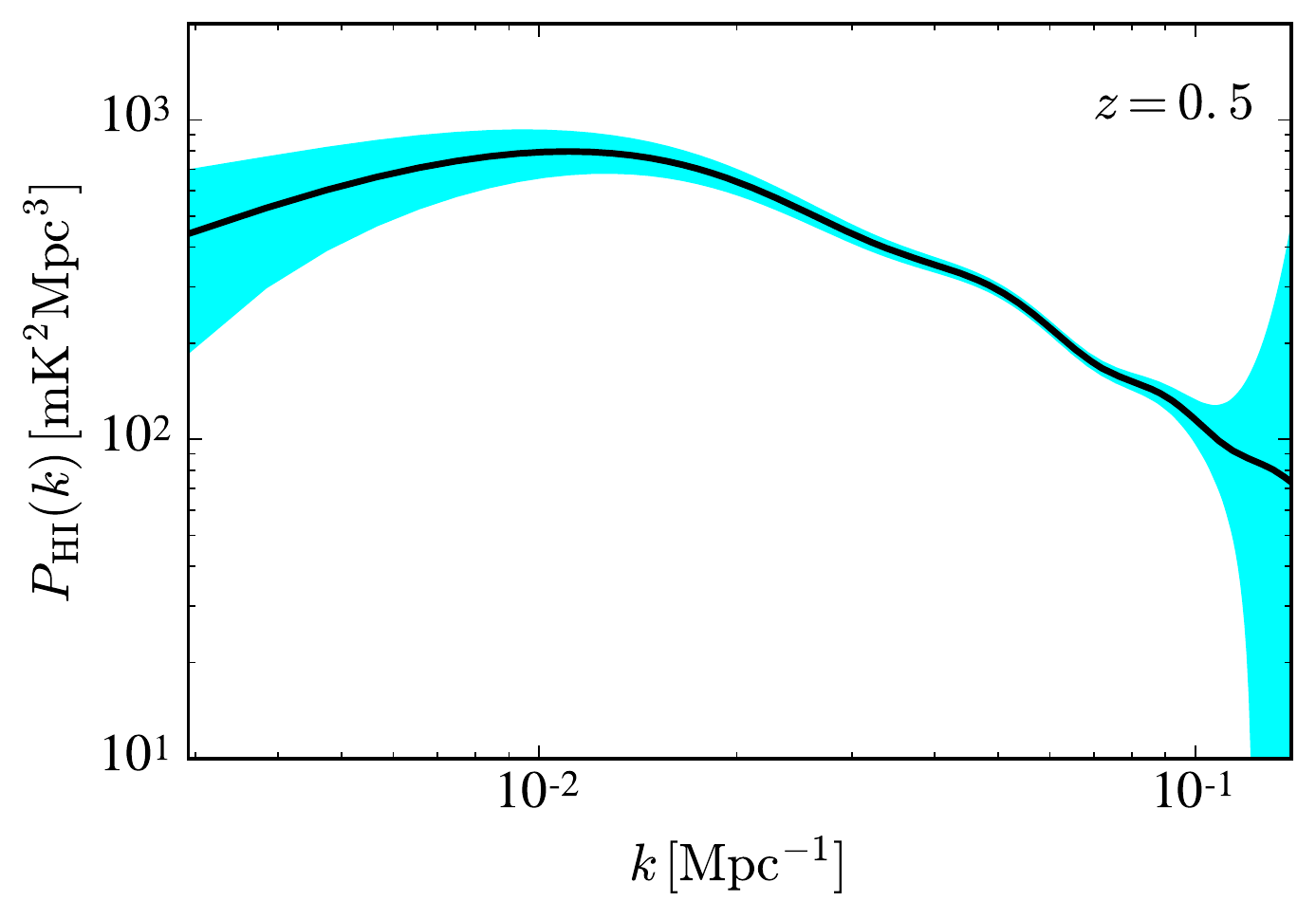}
\includegraphics[scale=0.5]{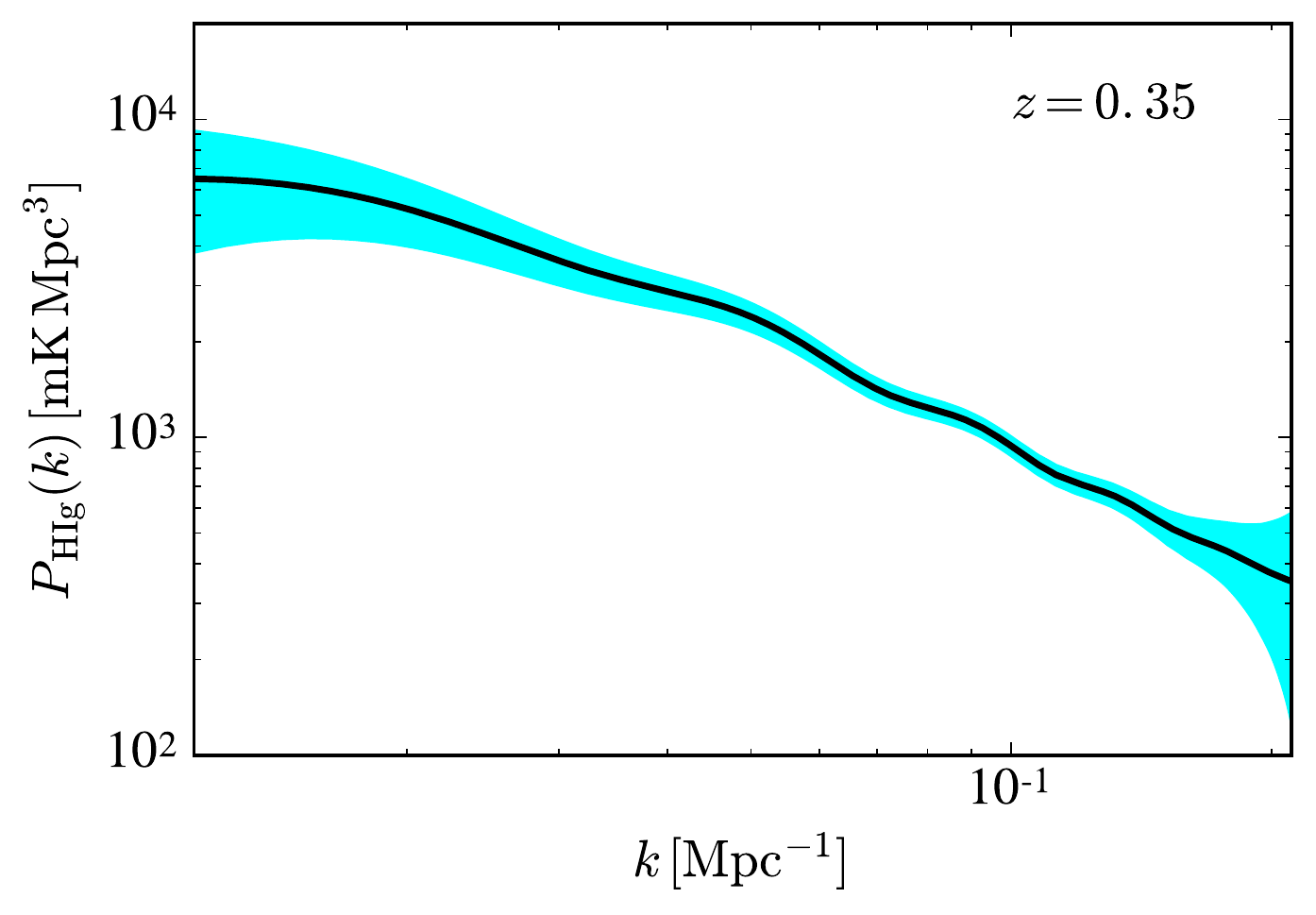}
\includegraphics[scale=0.5]{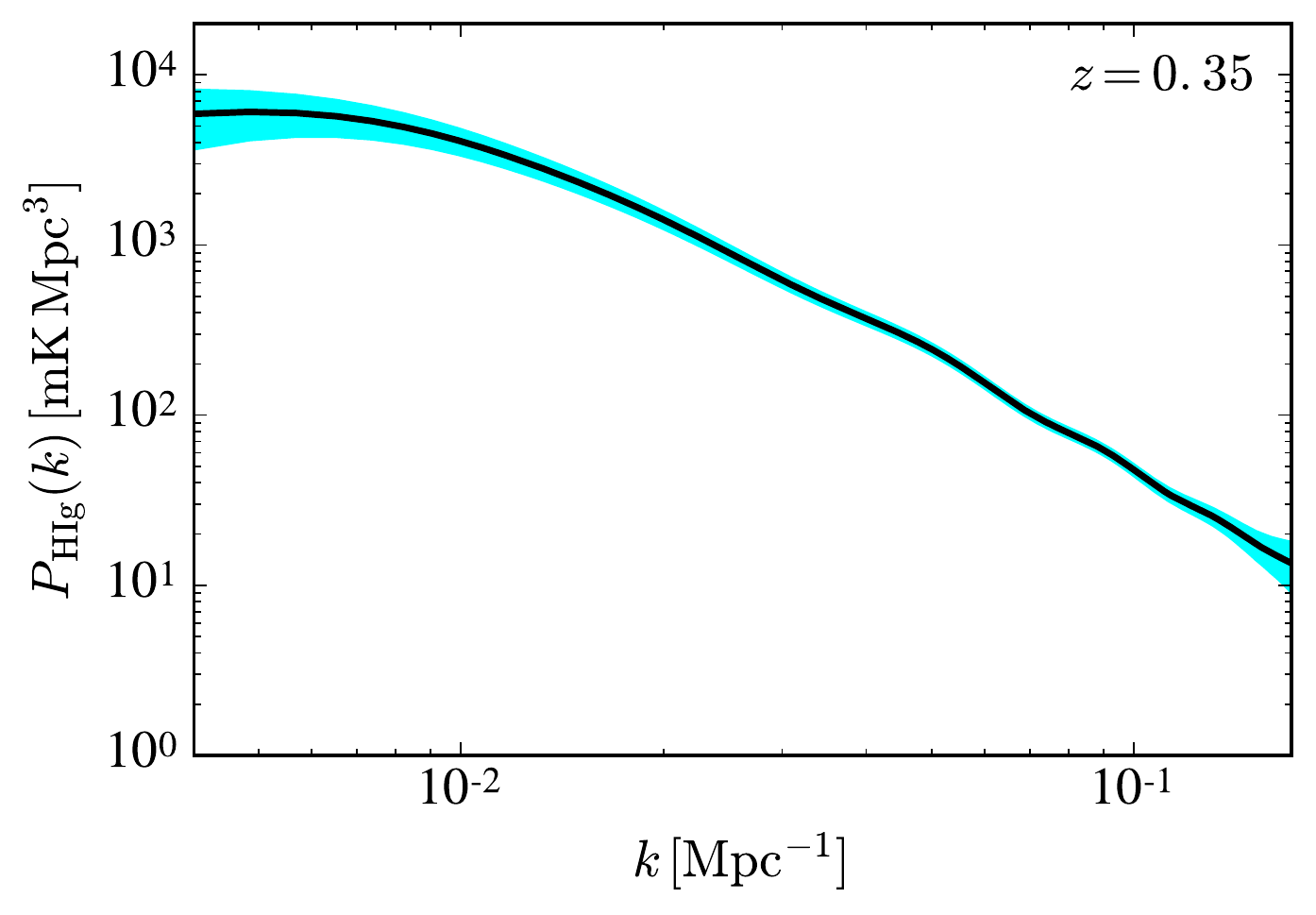}
\includegraphics[scale=0.5]{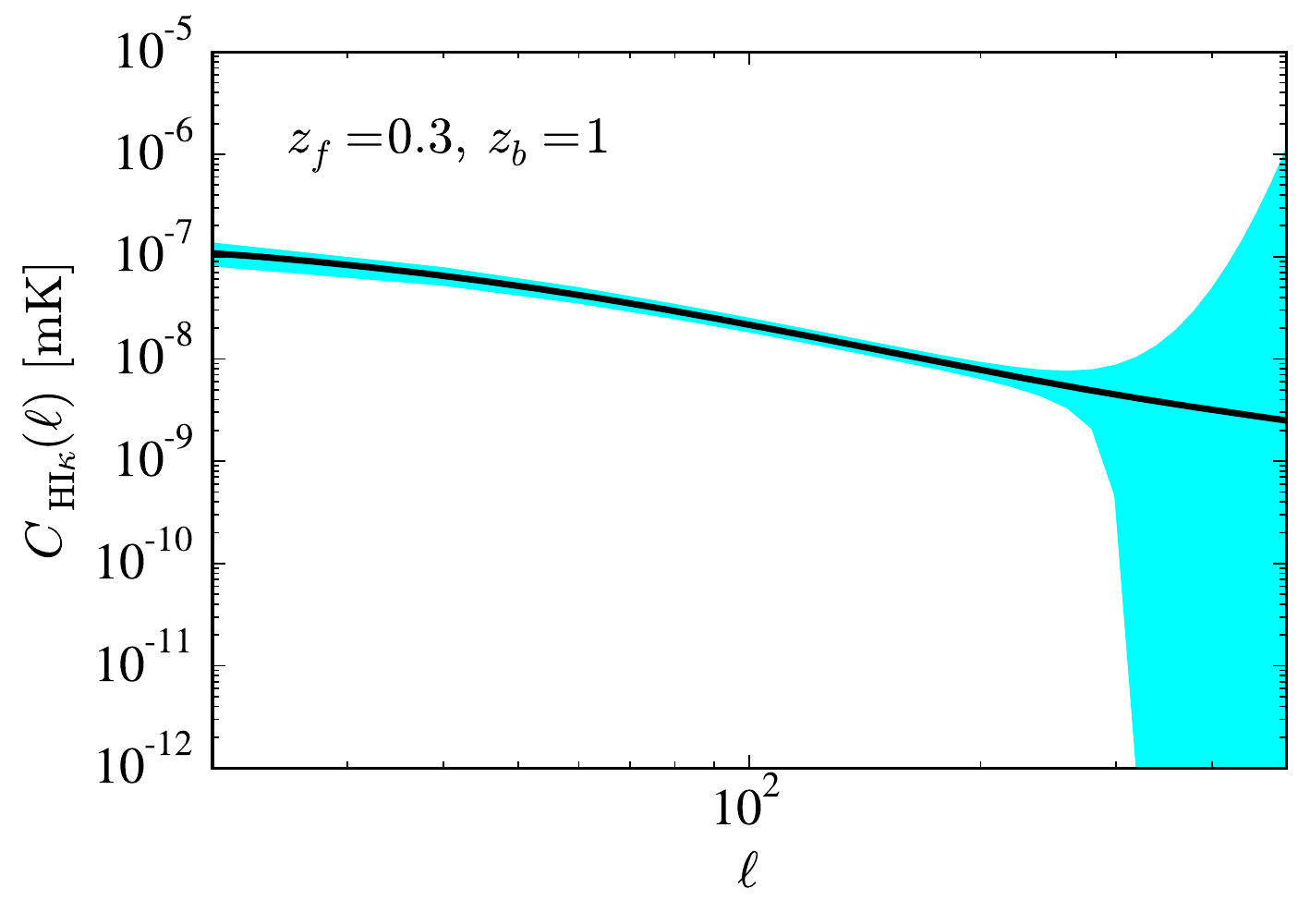}
\caption{HI detection with MeerKLASS, showing the expected signal power spectrum (black solid) and measurement errors (cyan). {\it Top left}: HI auto-correlation. {\it Top right}: Cross-correlation with  BOSS DR12-like  sample. {\it Bottom left}: Angle-averaged cross-correlation with DES-like  survey. {\it Bottom right}:  DES galaxies (at $z_b=1$) as background sources to detect the cross-correlation between the HI density field (at $z_f=0.3$) and lensing.}
\label{fig:PHI}
\end{figure}

After reionization, most HI will be found in dense systems inside galaxies. 
The total brightness temperature at a given redshift and in a unit direction ${\bf n}$ on the sky is 
\be\label{bt-rsd}
T_{b}(z,{\bf n}) \approx \overline{T}_{b}(z) \Big[1+b_{\rm HI}(z)\delta_m(z,{\bf n})-\frac{(1+z)}{H(z)}\,n^i\partial_i\big({\bf n}\cdot {\bf v} \big)\Big], 
\ee
where $b_{\rm HI}$ is the HI galaxy bias, $\delta_m$ is the matter density contrast, ${\bf v}$ is the peculiar velocity of emitters and the average signal $\overline{T}_{b}$ is determined by the comoving HI density fraction $\Omega_{\rm HI}$ (this is the proper HI density, divided by $(1+z)^3$ and by the critical density at redshift zero). The last term in the brackets is the usual RSD term. 
The signal will be completely specified once we have a prescription for the $\Omega_{\rm HI}$ and $b_{\rm HI}$. This can be obtained by making use of the halo mass function, ${dn}/{dM}$ 
and halo bias, relying on a model for the amount of HI mass in a dark matter halo of mass $M$, i.e. $M_{\rm HI}(M)$. We assumed a simple power law,
$
M_{\rm HI}(M) = A M^\alpha,
$
with $\alpha = 0.6$ and $A\sim 220$ both redshift independent and chosen to match the observed $\Omega_{\rm HI}$ at $z=0.8$  (see \citealt{2015aska.confE..19S} for details). In this paper, we used the following functions which fit the above prescription quite well:
\begin{eqnarray}
b_{\rm HI}(z) &=& \frac{b_{{\rm HI},0}}{0.677105} \big(6.6655 \times 10^{-1} + 1.7765 \times 10^{-1}\, z + 5.0223 \times 10^{-2}\,  z^2\big), \\
\Omega_{\rm HI}(z) &=& \frac{\Omega_{{\rm HI},0}}{0.000486} \big(4.8304\times 10^{-4} + 3.8856 \times 10^{-4}\, z - 6.5119 \times 10^{-5}\, z^2\big), \\
\overline{T}_b(z) &=& 5.5919 \times 10^{-2} + 2.3242 \times 10^{-1}\, z - 2.4136 \times 10^{-2}\, z^2\,\,{\rm mK}.
\end{eqnarray}


The first aim of the intensity mapping survey will be to measure the HI power spectrum, something that has not been achieved yet. In addition, we will take advantage of multi-wavelength coverage (e.g., BOSS, DES) to detect the signal in cross-correlation. This will reduce systematics and potentially make the detection even easier.
The MeerKLASS survey can be performed in the L-band ($0<z<0.58$) or  UHF-band ($0.4<z<1.45$). The noise properties of an intensity mapping survey like MeerKLASS have been described in detail in \citet{2015aska.confE..19S,Pourtsidou:2015mia}.


The HI power spectrum in redshift space follows from \eqref{bt-rsd}: 
\be\label{ps-rsd}
P^{\rm HI}(z,k,\mu)=\overline{T}_{b}(z)^2 b_{\rm HI}(z)^2 \big[1+\beta_{\rm HI}(z)\mu^2\big]^2P(z,k)  ,
\ee 
where $\mu = \hat{\bf k} \cdot {\bf n}$  and $\beta_{\rm HI}=f/b_{\rm HI}$ is the RSD parameter. 
The HI signal and measurement errors expected from MeerKLASS ($4000\, {\rm deg}^2$ sky area, $4000 \, {\rm hr}$ observation time) are shown in Fig.~\ref{fig:PHI}, top left, for a redshift bin of width $\Delta z=0.1$ centered at $z=0.5$. 
Here RSD have been neglected. 
Assuming a $\Lambda$CDM expansion history, the only unknown in $P^{\rm HI}$ is $\Omega_{\rm HI}b_{\rm HI}$. Employing a Fisher matrix analysis, we can calculate the expected constraints on $\Omega_{\rm HI}(z)b_{\rm HI}(z)$ \citep{Pourtsidou:2016dzn}, which are summarized in Table~\ref{tab:omHIbHI}. These are more than one order of magnitude better than current constraints from galaxy surveys, intensity mapping, and damped Lyman-$\alpha$ observations \citep{2015MNRAS.447.3745P}. 
\begin{table}
\begin{center}
\caption{\label{tab:omHIbHI} Forecast fractional uncertainties on $\Omega_{\rm HI}b_{\rm HI}$.}  
\begin{tabular}{ccc}
\hline
$z$ & $\delta(\Omega_{\rm HI}b_{\rm HI})/(\Omega_{\rm HI}b_{\rm HI})$ &\\
\hline
{\bf L-band} \\
0.1&0.010&\\
0.2&0.005&\\
0.3&0.005&\\
0.4&0.007&\\
0.5&0.009&\\
\\
\\
\\
\\
\hline
\end{tabular}
\begin{tabular}{ccc}
\hline
$z$ & $\delta(\Omega_{\rm HI}b_{\rm HI})/(\Omega_{\rm HI}b_{\rm HI})$ &\\
\hline
{\bf UHF-band} \\
0.6&0.011&\\
0.7&0.013&\\
0.8&0.015&\\
0.9&0.018&\\
1.0&0.022&\\
1.1&0.026&\\
1.2&0.030&\\
1.3&0.036&\\
1.4&0.042&\\
\hline
\end{tabular}
\end{center}
\end{table}

Detection can also be achieved by cross-correlating  with optical surveys \citep{2013ApJ...763L..20M}. The advantage is that systematic effects and foregrounds that are relevant for one survey but not for the other are expected to drop out in cross-correlation.
The cross-correlation power spectrum, neglecting RSD, is
$
P^{\rm HI,g}(z,k)=\overline{T}_{b}(z) b_{\rm HI}(z) b_{\rm g}(z) r P(z,k),
$ 
where the galaxy bias is modelled as $b_{\rm g}(z) = \sqrt{1+z}$, and $r$  accounts for the possible stochasticity between the tracers (expected to be close to unity on large, linear scales). 

Figure~\ref{fig:PHI}, top right, shows the expected signal and  errors for a BOSS DR12-like spectroscopic sample \citep{Satpathy:2016tct} at $0.2<z<0.5$. The assumed sky overlap  is 500\,deg$^2$ with 500\,hr MeerKAT time. 
The forecast constraint  is ${\delta(\Omega_{\rm HI}b_{\rm HI}r)}/{(\Omega_{\rm HI}b_{\rm HI}r)} \simeq 0.02$ \citep{Pourtsidou:2016dzn}.
Figure~\ref{fig:PHI}, bottom left, shows the expected (angle-averaged) signal and errors for a DES-like photometric sample \citep{2016MNRAS.460.1270D}, with 4000\,deg$^2$ sky overlap and 4000\,hr MeerKAT time. 
Finally,  Fig.~\ref{fig:PHI}, bottom right,  shows the expected cross-correlation signal and errors between the HI density field at $z_f=0.3$ and the lensing convergence $\kappa$ of background sources at $z_b=1$, using MeerKAT and DES (for details, see \citet{Pourtsidou:2015mia}).

\subsubsection{Baryon acoustic oscillations and redshift space distortions}

BAO  currently provide the most powerful probe of the Universe's expansion history ($D_A, H$), while RSD do the same for the rate of growth of structure ($f\sigma_8$). MeerKLASS aims to make the first detections of BAO and RSD in HI IM with enough signal to noise to constrain the cosmological model. It will in fact provide better constraints than any current survey (at any wavelength).
\begin{figure}
  \centering
  \includegraphics[width=0.95\linewidth]{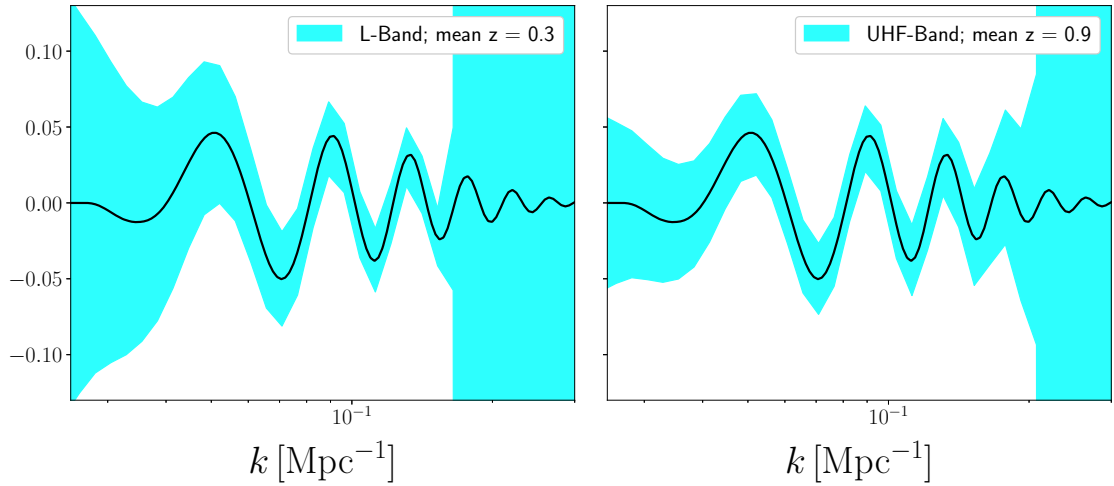}
\caption{BAO detection forecasts (4,000\,hr):  On larger scales, UHF performs slightly better (bigger volume at higher $z$ reduces cosmic variance). On smaller scales the nonlinear $k_\mathrm{max}$ cutoff limits the constraints. For the L-band $k_\mathrm{max} = 0.16$ Mpc${}^{-1}$ and for UHF $k_\mathrm{max} = 0.21$ Mpc${}^{-1}$, due to the higher redshift in UHF and the redshift scaling of $k_\mathrm{max}$. }
\label{fig:BAO}
\end{figure}
\begin{figure}
\centering
  \includegraphics[width=1.0\linewidth]{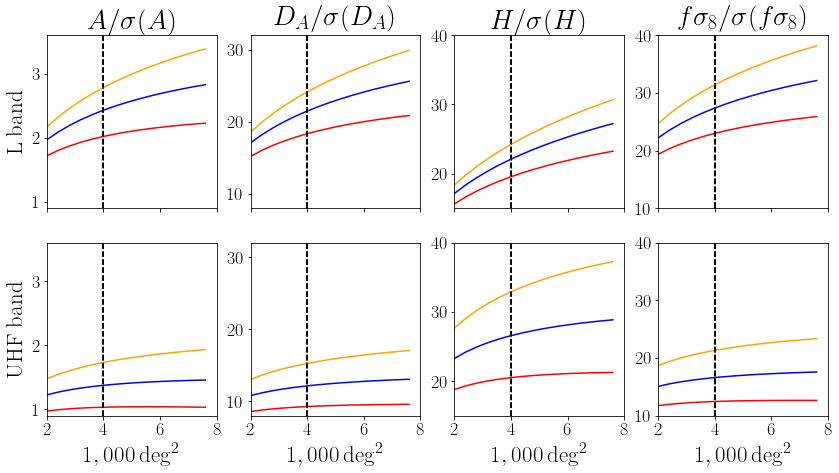}
\caption{Signal to noise as a function of survey area. Columns from left to right show: BAO amplitude parameter $A$, angular diameter distance $D_A$, Hubble rate $H$, growth rate $f\sigma_8$. Curves are for: 2,000 (red), 4,000 (blue), 8,000\,hr (orange). Upper row: $z=0.4$ in L-band, lower row: $z=1.0$ in UHF-band (redshift bin size = 0.1). Vertical dashed lines indicate 4,000\,deg$^2$. The L-band outperforms UHF for all observables except $H$.}
\label{fig:HDAfs8A_z04}
\end{figure}


Figure \ref{fig:BAO} shows the forecast constraints on the BAO wiggles in the power spectrum, for  L- and UHF-band (combining all the redshift information). Figure \ref{fig:HDAfs8A_z04} shows the expected signal to noise from  BAO and RSD for different parameters, as a function of sky area and for a few observation times, at $z=0.4$ and $z=1.0$ (with a z bin of 0.1). The forecasting procedure follows the approach described in \citet{2014arXiv1405.1452B} (see also \citealt{Pourtsidou:2016dzn}).
Looking in particular at the constraints on $A$, the overall BAO amplitude (left), we see that the signal to noise is better in the L-band. Moreover, 4,000\,hr and 4,000\,deg$^2$ is seen as a good balance. For the same 4,000\,hr, more sky area will be better, but not so much better as to offset the penalty on the other science, such as the continuum science case. On the other hand, 2,000\,hr provides a S/N below 2 which makes it a risky option in terms of making a clean signal detection.
Note also that we want to use sky areas with good multi-wavelength coverage, which will essentially set the maximum available area to about 6,000\,deg$^2$ for the next six years.
The upper right panel in Fig. \ref{fig:HDAfs8A_z04} shows that detection of the growth rate with MeerKLASS will be possible with a signal to noise of almost 30 in the L-band. Note that this analysis does not take into account the foreground contamination, which will probably make higher signal to noise maps more preferable.




\subsubsection{Multi-tracer constraints on primordial non-Gaussianity}

\begin{figure}
	\centering
	\includegraphics[width=0.48\textwidth]{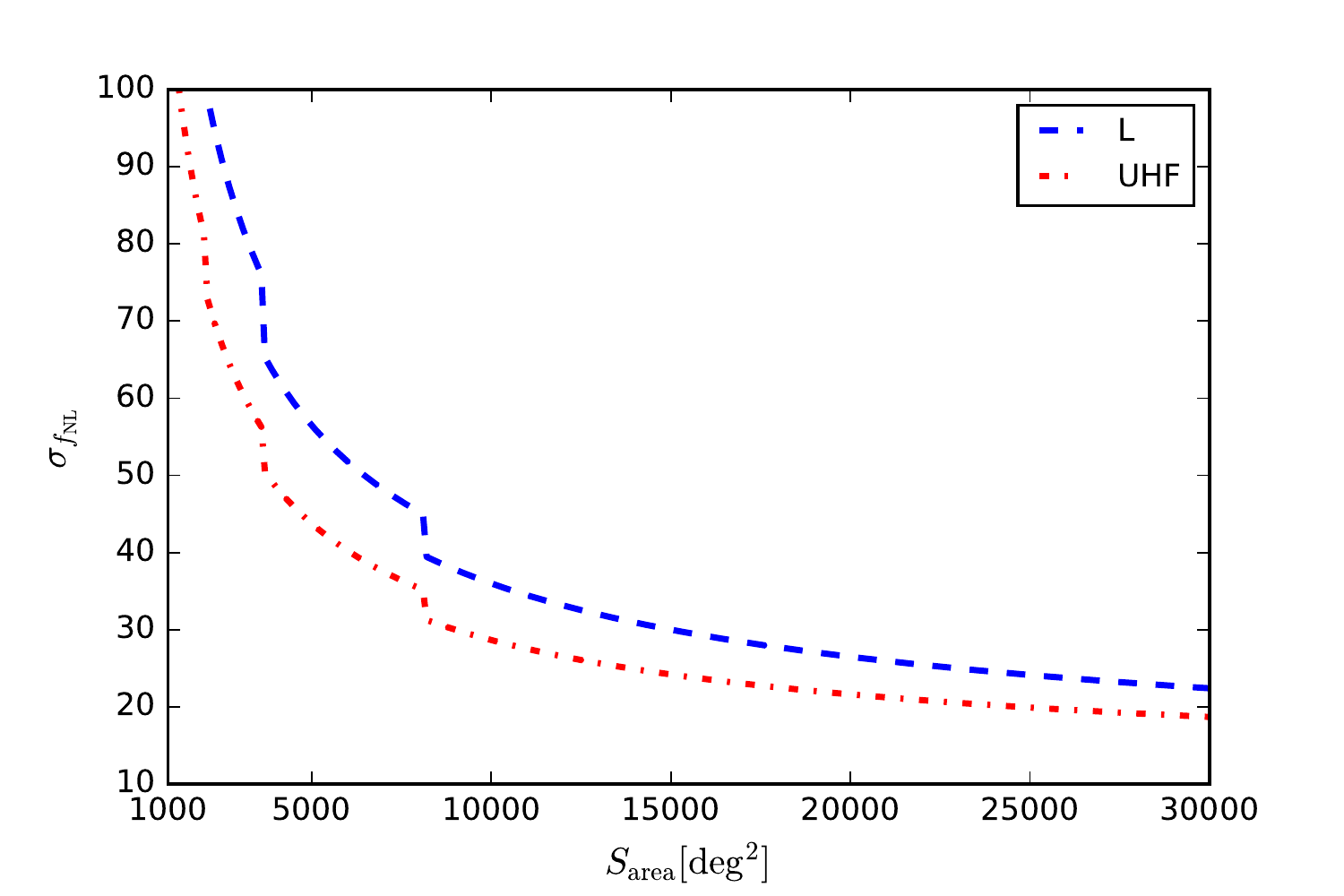}\ \ 
	\includegraphics[width=0.48\textwidth]{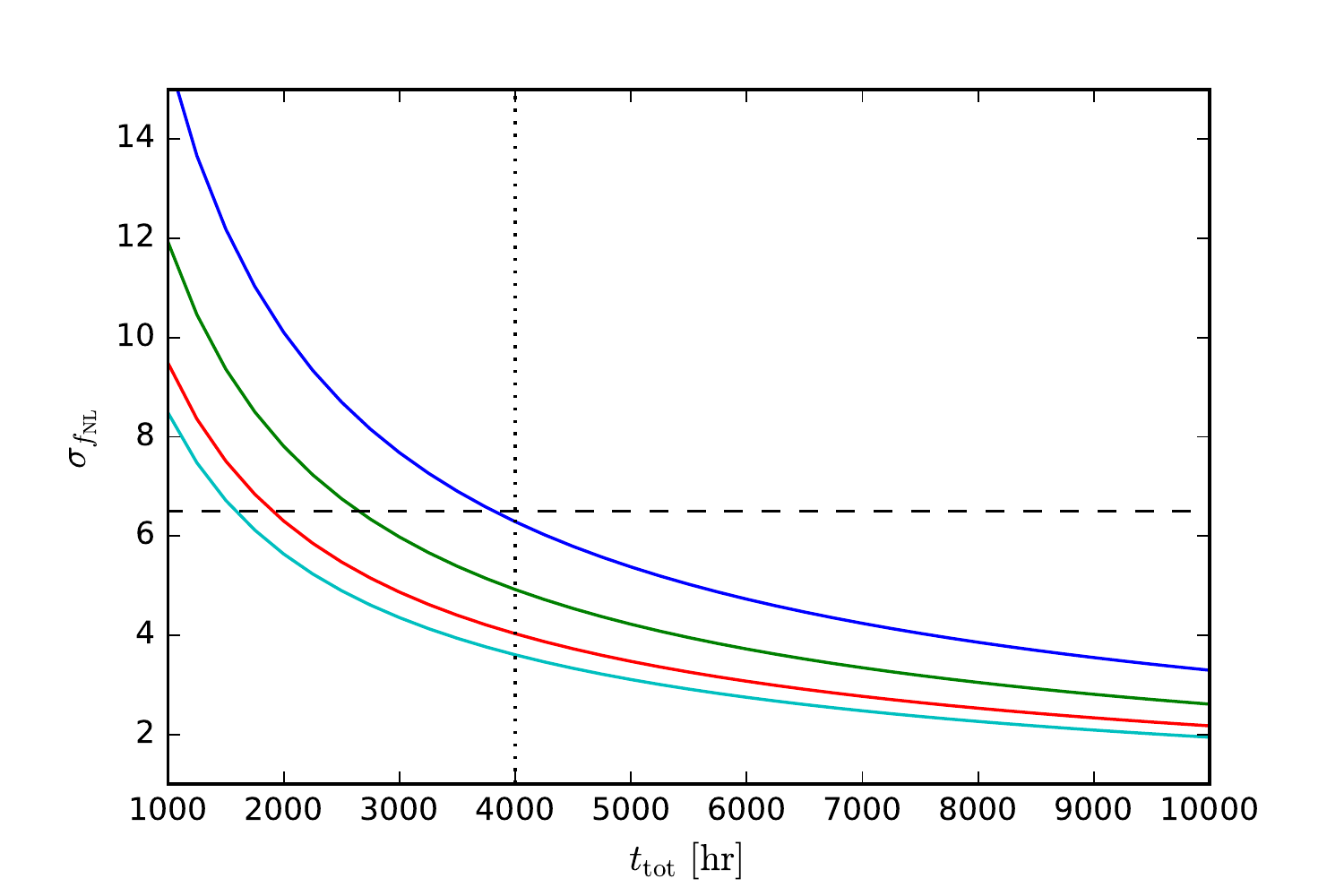}
	\includegraphics[width=0.48\textwidth]{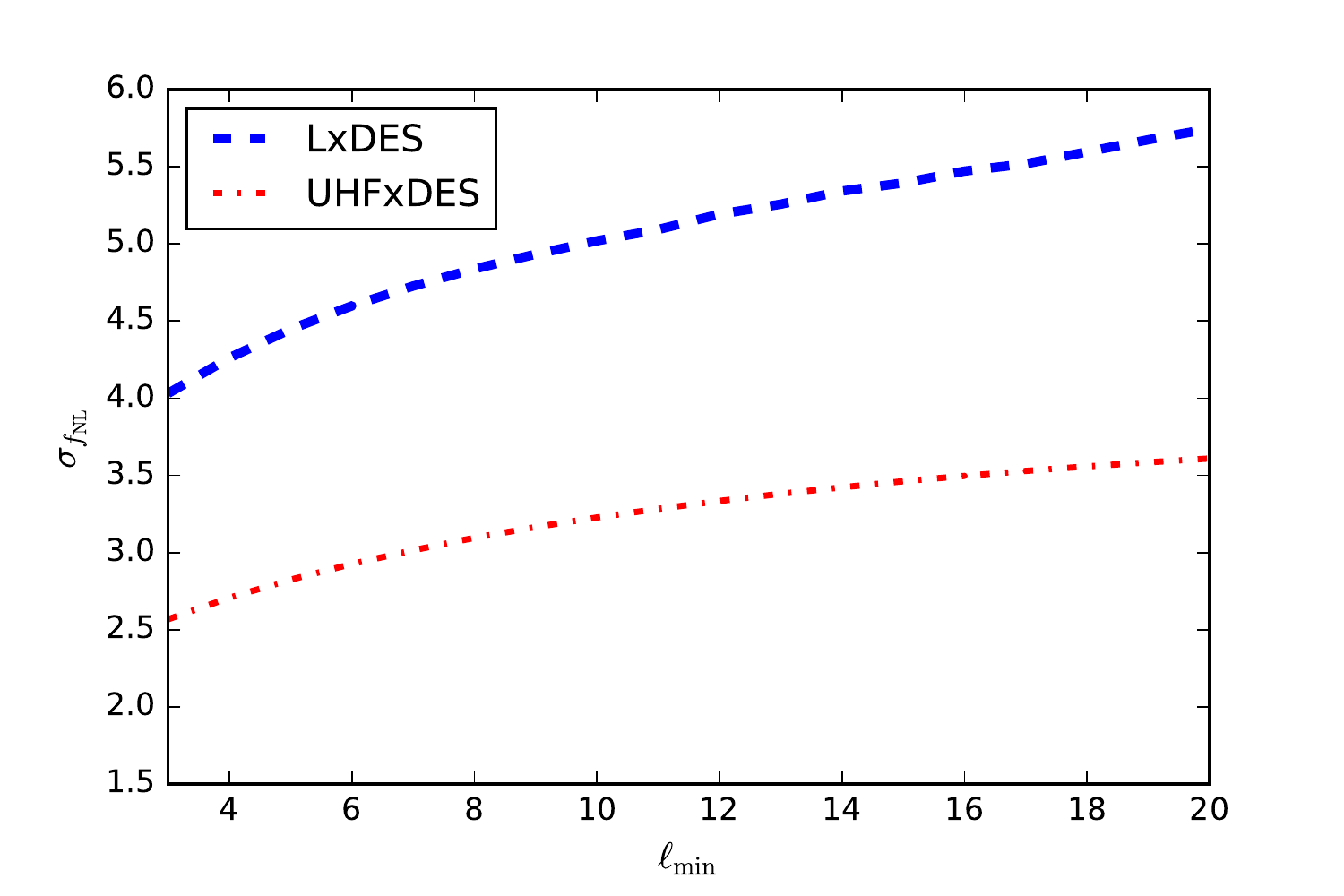}\ \ 
	\includegraphics[width=0.48\textwidth]{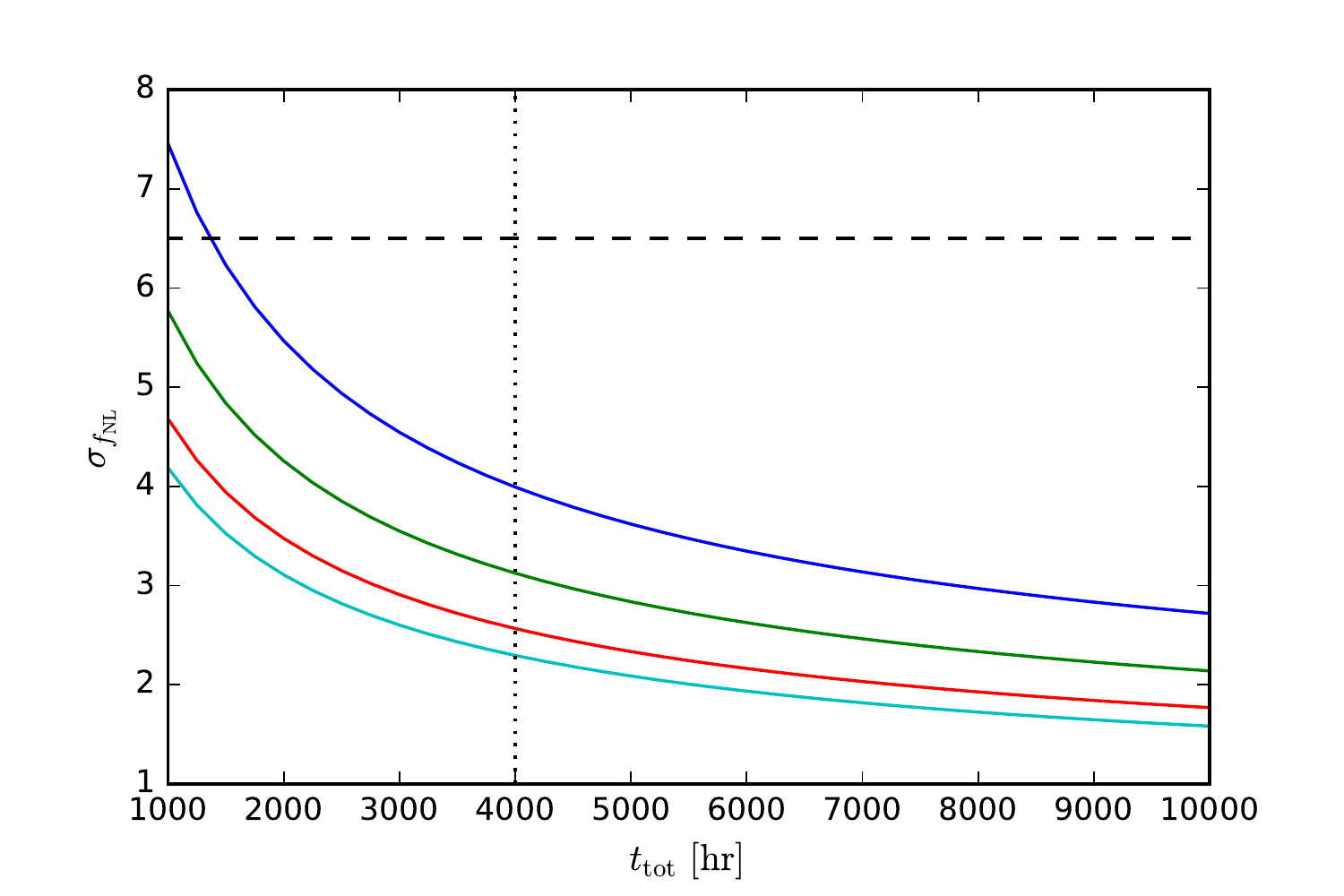}
	\caption{Error on $f_{\rm NL}$. {\em Top Left}: MeerKAT L-band and UHF-band, against sky area, for 4000\,hr of observation. {\em Top Right:} Using multi-tracer with MeerKAT L-band and DES, against observation time. Solid lines correspond to overlap areas (from top to bottom): 2000, 3000, 4000, 5000\,deg$^2$. Vertical dotted line at 4000\,hr indicates the proposed MeerKLASS observation time. Horizontal dashed line indicates the Planck constraint $\sigma(f_{\rm NL})=6.5$. {\em Bottom Left}: Using multi-tracer with MeerKAT L-band or UHF-band and DES, against size of an individual patch ($\ell_{\rm min}$) for a fixed total survey area of 4000\,deg$^2$ and 4000\,hr of observation. {\em Bottom Right:} Same as Top Right but using multi-tracer with MeerKAT UHF-band and DES.}
	\label{fig:fnlMK}
\end{figure}

In recent years we have seen an increased interest in testing the Gaussianity of primordial perturbations, motivated by the plethora of inflationary models that have appeared since the first inflation model was proposed. The simplest models predict that perturbations are very close to Gaussian while non-standard models or models with isocurvature fields will naturally source non-linear corrections. Primordial non-Gaussianity (PNG) leaves a `frozen' imprint in the power spectrum on very large scales. Measuring this signal provides a powerful probe of inflation models. The local type of PNG has amplitude $f_{\rm NL}$ and introduces a scale-dependent clustering bias. On super-equality scales  \citep{Dalal:2007cu,Matarrese:2008nc}:
\bea\label{bng}
 b(z,k)=b_{\rm G}(z)+ \big[ b_{\rm G}(z) -1\big]\alpha(z)\,f_{\rm NL}\frac{H_0^2}{k^2}
\eea
where $b_{\rm G}$ is the linear Gaussian bias and $\alpha$ is a known function. The PNG signal thus grows on the largest scales. For a single tracer of the dark matter distribution, this signal is eroded by cosmic variance, and even the next-generation ultra-large survey volumes are unable to achieve $\sigma(f_{\rm NL})<1$ \citep{2013PhRvL.111q1302C,2015aska.confE..25C,Alonso:2015uua}.  Figure \ref{fig:fnlMK} (top left) shows the constraints achievable with MeerKAT on its own, a lot weaker than the current state of the art delivered by the Planck experiment, $\sigma(f_{\rm NL})=6.5$.

Cosmic variance can be beaten down using multiple tracers \citep{2009PhRvL.102b1302S,2014MNRAS.442.2511F,2015ApJ...812L..22F}, and this is applied to MeerKAT and DES in \citet{Fonseca:2016xvi}. For more details on the multi-tracer technique look at the references therein. The results for L-band are presented in Fig.~\ref{fig:fnlMK} (top right). The multi-tracer technique can achieve errors on $f_{\rm NL}$ better than those of Planck (horizontal dashed line), even with a conservative assumption on the overlap sky area between MeerKAT and DES (3000 or 4000\,deg$^2$), and with a maximum of 4000\,hr integration time. MeerKLASS can therefore deliver the best constraints on $f_{\rm NL}$ before SKA, Euclid and LSST deliver data. These constrains would be tighter using the UHF-band, as one can see in Fig.~\ref{fig:fnlMK} (bottom right), although the effect of foregrounds could affect this.

The last thing to note is that the forecasted error on $f_{\rm NL}$ does not degrade rapidly if one breaks down the total survey area into smaller patches of the sky. This is visible on the bottom left panel of Fig.~\ref{fig:fnlMK}. The sole effect of considering smaller non-contiguous patches lies on the minimum $\ell$ available. Although this option would be suboptimal, one can still deliver a measurement of $f_{\rm NL}$ improving on Planck results even for 4 patches of 1000 $\deg^2$ or 8 of 500 $\deg^2$.

\subsubsection{Cross correlations with the CMB}

Direct correlations of the HI intensity field with the CMB are challenging because CMB observables, such as the lensed field or the Sunyaev-Zel'dovich effect, are line-of-sight projected quantities with broad redshift support, and it is precisely these long-wavelength radial modes that are lost in the 21cm intensity field when removing galactic foregrounds. However, a cross-correlation signal can be recovered using higher order correlations of the HI density field \citep{Moodley:2017} or tidal field reconstruction \citep{2016arXiv161007062Z}. Cross-correlations with the reconstructed CMB lensing field \citep{Moodley:2017} will allow us to constrain the high-redshift matter power spectrum and the change in the HI density and bias with cosmic time.

\subsection{Continuum galaxy survey: constraining dark energy}

A continuum survey can be done simultaneously with an HI intensity mapping survey. Contrary to line surveys, requiring high resolution in frequency, these surveys integrate over a continuous interval in frequency ($\sim 300\,$MHz), providing high sensitivity and the detection of a large number of radio galaxies ($>10^8$ with SKA1). Unlike wide-field optical surveys, radio continuum emission is unaffected by dust, and AGN and galaxies will be detected to high redshifts. 
MeerKLASS will be able to detect almost 10 million radio galaxies in L-band (a factor of 10 more than current surveys), which will allow us to impose stringent constraints on dark energy.
\begin{figure}[h]
	\centering
	\includegraphics[width=0.5\textwidth]{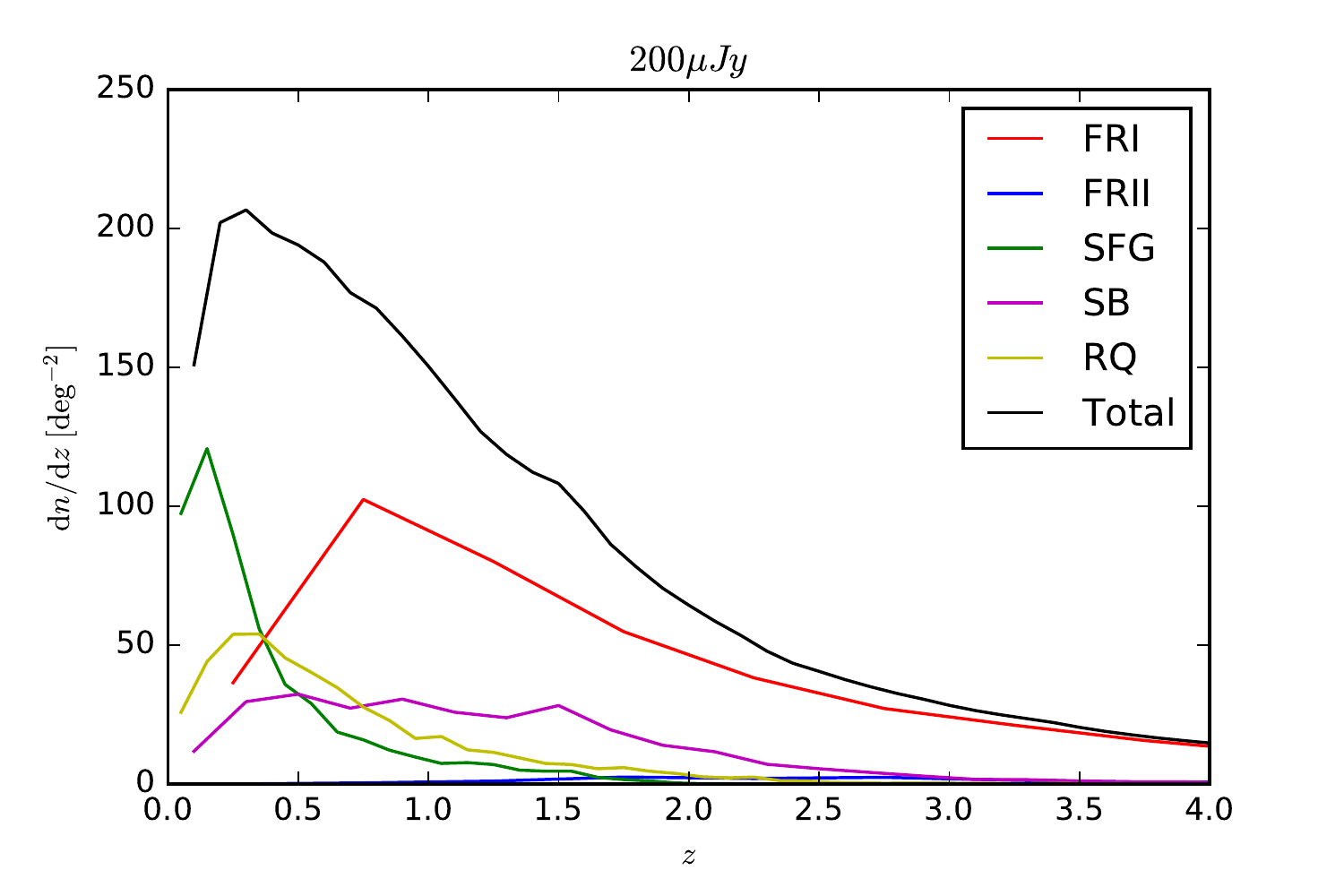}\includegraphics[width=0.5\textwidth]{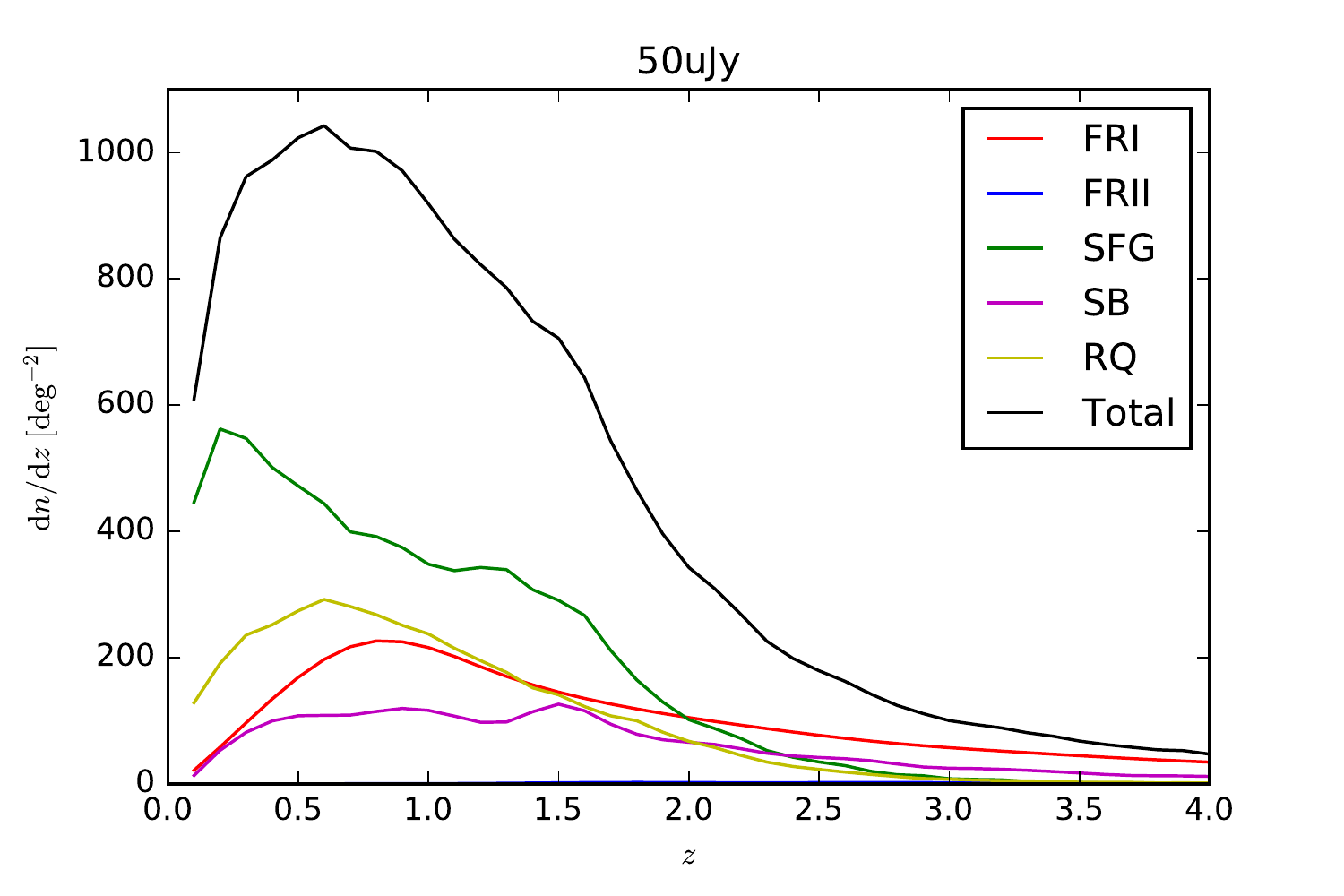}
	\caption{Redshift distribution of continuum sources detected at 200 ({\em left}) or $50\,\mu\mathrm{Jy}$ ({\em right}, the MeerKLASS flux cut) for various radio galaxy populations (different colours) and the total (black). Most MeerKLASS galaxies will be star forming.}\label{fig:dNdz}
\end{figure}

Since continuum radio surveys reach higher redshifts ($z\gtrsim2.5$) than  optical/near-IR counterparts, they are in principle ideal  to track the evolution of dark energy.  Figure~\ref{fig:dNdz} shows the redshift distribution per square degree of various radio galaxy populations (coloured curves) that will be detected by a continuum survey with a flux cut at 200 (left) or 50\,$\mu$Jy (right). The latter is the expected $\sim10\sigma$ level for MeerKLASS, while the former roughly corresponds to the expected $\sim10\sigma$ level for the planned ASKAP EMU full sky survey (this could be more than $\sim10\sigma$ depending on EMU's final specs).

\begin{figure}[h]
	\centering
	\includegraphics[width=10cm, height=8cm]{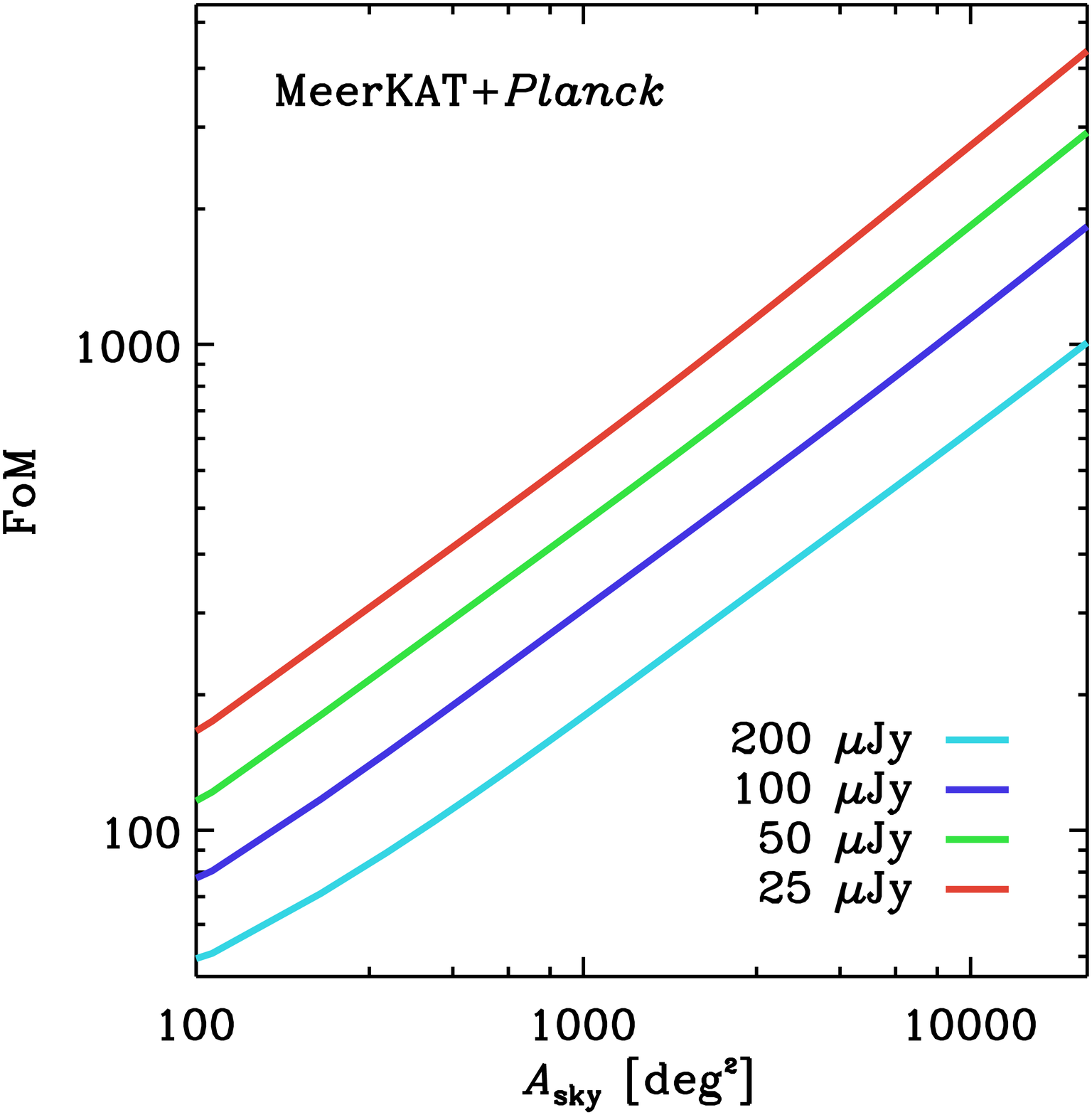}\ \ 
	\caption{\label{fig:FoM_Adeg2}Dark energy equation-of-state figure of merit against sky area for MeerKAT with different flux cuts (including a prior from {Planck}).}
\end{figure}

The distributions have a high-$z$ tail, with non-negligible source number densities as deep as $z\sim4$. However, this major advantage of radio continuum for cosmology is spoiled by the featureless synchrotron spectrum, which does not allow for the measurement of redshift. To overcome this, we use redshift information from cross-identifying sources via observations at different wavelengths \citep{2012MNRAS.427.2079C}. We use 5 photometric bins in  $0.01<z<0.8$ and then a large bin containing all the remaining high-$z$ galaxies. Then we forecast the precision of a continuum angular power spectrum  in constraining the dark energy equation-of-state parameters $w_0$ and $w_a$, representing the present-day value and the evolution with redshift. Figure~\ref{fig:FoM_Adeg2} shows the dark energy figure of merit, with impressive results for MeerKLASS.
\\
\\

\subsection{Dark matter detections}
MeerKLASS can be fruitfully employed for dark matter (DM) searches in both dwarf galaxies (dSphs) and in galaxy clusters using its radio signals from annihilation, as discussed in recent works \citep{2014JCAP...10..016R, 2015aska.confE.100C, 2016JCAP...05..013B}. Specific analysis of  the most promising dSphs found in the MeerKLASS survey down to 5 uJy in L-band are probably the best strategy to pursue this. With a few hours integration time over the region of a selected single dSph,  one can increase the sensitivity by a factor of  about 50 with respect to the present observations, thus largely improving the ability to set constraints (or possibly detect) DM indirect radio signals (see Fig.6 in \citealt{2014JCAP...10..016R}). The analysis of DM annihilation radio signals in galaxy clusters will be done out of the cluster blind survey, focusing on those systems which show the best spatial offsets between DM and baryonic matter, as to facilitate the analysis of possible DM indirect radio signals.

\section{Radio continuum: Galaxy evolution}

It is now accepted that most massive galaxies contain a central supermassive black hole that can undergo periods of accretion to create an Active Galactic Nuclei (AGN).  
These AGN show a strong interplay with the star formation activity of galaxies and play a significant role in the their evolution. This is evident from comparisons made between the AGN and star formation rate density of the Universe which has been shown to follow each other over the course of cosmic history \citep{Franceschini99}. More locally, the bulge masses of galaxies show a correlation with black hole masses, which suggests that an AGN feedback mechanism exists \citep{Kormendy1995,Magorrian98}. Incorporating AGN feedback into semi-analytical galaxy simulations \citep{Bower2006, Croton2006} has also been used as a mechanism to quench star formation, which produces the observed bimodality in the colours of galaxies \cite{Strateva2001,Baldry2004}.

\subsection{AGN}

Radio continuum emission at  $\sim 1.4$\,GHz is dominated by synchrotron emission from AGN and star forming (SF) galaxies, and provides a clear view of the emission mechanisms, free of dust obscuration.
The depth and area of MeerKLASS is ideally suited for investigating evolution of AGN and the impact they have on their host galaxy and wider environment.
%
The most powerful AGN ($L_{\rm 1.4 \,GHz} > 10^{25}\, \rm{W Hz}^{-1}$) evolve strongly over cosmic time, with number density  increasing by a factor of $\sim 1,000$ from $z = 0$ to $2$, 
and then declining beyond $z \sim 3$.
For lower luminosity AGN, the picture is more uncertain, with some studies finding little or no evolution \citep[e.g.][]{McAlpine2013, Prescott2016}
and others finding more significant evolution \citep[e.g.][]{Sadler2007}. 
Lower luminosity AGN dominate the  AGN feedback required by the latest simulations of galaxy evolution, and understanding their evolution as a function of accretion rate and environment is crucial for our general understanding of massive galaxy evolution.

MeerKLASS will provide the  data necessary to answer this question, particularly when combined with the wealth of multi-wavelength data available over the proposed fields.  Using the S$^3$ simulation 
as a guide, we will discover  $\sim 50,000$ FRI-type galaxies at $z<1$ and measure the evolution out to $z\sim 2$, where it has yet to be measured with any precision. Furthermore, using optical and near-IR data (DES, KIDS, VIKING, VHS), we will  relate the radio emission to the host galaxy properties and the accretion rate of the AGN, using mid-infrared data from WISE and the next generation of multi-object spectrographs.

\subsection{High-redshift AGN}

The large sky area of MeerKLASS makes it ideal to search for rare, high-$z$ AGN. These can be used to probe reionization, and can help constrain models of structure formation in the $\sim 1$\,Gyr period following recombination. LOFAR and MWA hope to detect the 21cm transition line at $z \gtrsim 6$, predominantly by searching for signatures via statistical fluctuations in the power spectrum. 
However, it is also possible to study reionization via 21cm absorption, if one can identify sufficiently distant background sources. This is one of the key science questions for the SKA.

Many attempts have been made to find $z > 6$ radio sources, mainly using `steep-spectrum' selection, 
with the highest redshift radio galaxy at $z = 5.19$. 
The major limitation is the difficulty in obtaining complementary imaging data in the near-IR, which are needed to reliably eliminate low-$z$ sources. The advent of wide-field near-IR surveys (Spitzer, UKIRT, VISTA, WISE) means these data are now available over sufficiently large sky areas to filter out sources at $z < 2$, and leave the remaining non-detections for spectroscopic follow-up \citep[e.g.][]{2009MNRAS.398L..83J}. 

With the proposed survey area, resolution and depth (required to reliably identify infrared counterparts), we expect  to find, in combination with VIKING, about ten $z > 6$ radio sources with strong emission lines, plus another $\sim 300$ lower-luminosity sources which should still be bright enough for SKA 21cm absorption studies.




\subsection{Star-forming galaxies}
\label{sec:SFGalaxies}

The star formation history of the Universe is of great importance to test cosmological theories and also for many applications in  astrophysics, such as estimating the build-up of stellar mass in galaxies, constraining the initial mass function, testing models of chemical evolution or constraining the nature of stellar deaths in supernovae and gamma-ray bursts. 
It is also one of the key science cases to be addressed by the SKA \citep{Jarvis2015_SF}.
%
Radio emission from SF galaxies is produced from free-free emission in HII regions near massive stars and cosmic ray electrons produced by supernova remnants, and offers an unobscured probe of the Universe without the need for extinction corrections. 

A survey like MeerKLASS, containing millions of galaxies, will constrain the luminosity functions of AGN and SF galaxies to much higher accuracy, and down to much lower luminosities than previously. Combining the radio data with both spectroscopic and photometric redshifts from optical and near-IR surveys will enable the investigation of how this evolution depends on stellar mass and environment. These are critical for relating the observations to the simulations and thus making important insights into the physics of star formation and feedback processes. 


\subsection{Clustering measurements and AGN environments}

The clustering properties (e.g. bias) of radio sources can be used to provide important constraints on how galaxy formation and evolution are related to the underlying dark matter distribution. 
MeerKLASS will allow us to extend this work several orders of magnitude deeper, and therefore probe the clustering of faint (sub-mJy) radio sources for the first time. 

Clustering measurements can be used to determine the fraction of dark matter haloes of a given mass which host radio-loud AGN (FRIs or FRIIs for example) and also what fraction of these exhibit significant star formation activity, providing the information to measure the duty cycle 
of the AGN activity, which can be used to estimate the timescale of activity and how it relates to the lifetime of star-formation activity in host galaxies. Constraining the duty cycle of AGN, and how this is affected by accretion mode and morphology, is vital to understanding AGN feedback. Furthermore, novel use of cross-correlations between quiescent and active galaxies can be used to investigate how activity depends on halo mass and possibly the age of the dark matter haloes. 

The proposed survey will take studies of the clustering of radio sources to the next level: with the large area covered and breadth of ancillary data, we will be able to determine the evolution of the clustering of radio sources as a function of their morphology, luminosity and accretion mode up to $z \sim 2$.


\section{Clusters}

MeerKLASS will observe total intensity and polarization over 4,000\,deg$^2$ down to a flux limit of $\sim5\,\mu$Jy/beam in L-band. More than 1,500 galaxy clusters above a $M_{500}$ mass limit of $3 \times 10^{14} M_\odot$ are expected within the survey area (assuming $\sigma_8 = 0.80$). This blind survey will produce sensitive radio observations of a statistically significant number of galaxy clusters above this mass limit and over a broad redshift range, from $z=0$ up to $0.5-0.8$, depending on the evolution of the physical mechanisms producing the diffuse radio emission.

\subsection{Diffuse cluster emission}
Most diffuse emission studies have been carried out on high-mass ($M_{500} > 6 \times 10^{14} M_\odot$), low redshift ($0.2 < z < 0.4$) systems due to the faint nature of the emission and sensitivity limits of current instruments. The MeerKLASS survey is expected to detect a radio halo and/or relic in approximately 200 clusters, lower in mass and over a broad range of redshift, greatly extending the current number of $\sim$ 65 sources \citep{2015ApJ...813...77Y}. This statistically significant sample of diffuse emission is crucial for advancing the understanding of its formation and evolution.

MeerKLASS will measure fluxes and spectral indices across the broad MeerKAT band for each diffuse source (as well as wide band spectral indices through the correlation with other radio surveys, e.g. NVSS, SUMSS, TGSS, where there is overlap). For radio halos, this will allow a separation of the detected population based on their formation theories: re-acceleration during cluster mergers creates turbulent halos, whereas off-state halos are found in less disturbed systems and are generated through hadronic processes. 
The sensitivity of MeerKAT also allows for an investigation into the evolution and formation of radio halos by studying the rate of occurrence in low mass and/or high redshift systems. Polarization data will be crucial to test the formation models for shock-induced radio relics through magnetic field orientations and inferred Mach numbers. The blind survey will also detect anomalous diffuse sources which do not conform to the current emission characterizations, which will further the understanding of the evolution and properties of diffuse cluster emission. Finally, with the 10 arcsecond L-band resolution of MeerKAT, a statistically significant population of mini-halos should be detected around low-redshift BCGs. This population will be used to begin to investigate whether mini-halos are inherently separate phenomena from radio halos or instead part of the latter's evolutionary process. 

MeeKLASS will also be able to detect and characterize many radio relics within and around galaxy clusters thus shedding light on their origin (re-acceleration of seed electron or connection with radio galaxy remnants). In this context the survey will allow to measure radio-derived Mach numbers of relics and extend and characterize the P$_{1.4}$-Mach number correlation which has been proposed as a powerful test of radio relic origin \citep{2017MNRAS.471.4747C}.

\subsection{Star formation in clusters}

As noted in Section~\ref{sec:SFGalaxies}, MeerKLASS will detect millions of star forming galaxies. The combination of MeerKLASS with large, Sunyaev-Zel'dovich (SZ) effect selected galaxy cluster surveys, such as that being conducted by the Advanced Atacama Cosmology Telescope \citep[AdvancedACT;][]{DeBernardis_2016}, will allow us to investigate the evolution of star formation in and around clusters, over half of the age of the Universe. The sensitivity of MeerKLASS ($\sim 5 \mu$\,Jy RMS) will allow individual galaxies with star formation rates (SFR) $ > 20$\,M$_{\odot}$/yr at $z < 0.7$ to be detected at 5$\sigma$, assuming the \citet{Bell_2003} relation between SFR and 1.4 GHz\,luminosity. We will measure how the star formation rate per unit cluster mass evolves, and investigate how this varies with cluster mass. The clean, SZ selection of the clusters in the MeerKLASS field will simplify the interpretation of the results, by easing comparison with cosmological simulations. The sensitivity reached by the MeerKLASS observations is comparable to surveys of $\sim 10-100$s of clusters over a similar redshift range that have been conducted at 24\,$\mu$m with MIPS on the Spitzer Space Telescope \citep{Webb_2013}, or with Herschel at 250\,$\mu$m \citep{Alberts_2014}. The advantage of our survey will be an increase in the sample size by 1-2 orders of magnitude, and reliable masses based on SZ observables (previous surveys are based on IR-selected cluster samples).

\subsection{Cluster magnetic fields}
The origin of cosmic magnetic fields is still an open field in terms of a full understanding of the processes which generate or amplify them. One of the methods to probe the strength and orientation of magnetic fields in the ICM, necessary for understanding cosmic magnetism, is through Faraday rotation of background sources. With its 5\,$\mu$Jy/beam sensitivity limit, MeerKLASS is expected to detect $\sim25$ polarized sources per square degree \citep{2014ApJ...785...45R}, several of which will be in or behind galaxy clusters. Magnetic field studies in clusters over a wide range of mass and redshift will provide crucial statistical information on the properties and evolution of magnetism on cluster scales.

\section{MeerKLASS as an extragalactic HI survey}

MeerKLASS promises to resolve a statistically significant number of galaxies over the widest range of galaxy \ion{H}{i} masses, while at the same time maintaining a high sensitivity for low density extended structures ($4\cdot 10^{19}\,{\rm atoms}\,{\rm cm^{-2}}$ at $4\,\sigma_{\rm rms}$ over $20\,\rm km \, s^{-1}$ at $HBPW\,=\,30^{\prime\prime}$). It therefore perfectly complements WALLABY on ASKAP \citep{2012PASA...29..359K}, which will be unable to delve into the low-column density regime, as traced with MeerKLASS, and may be affected by confusion. This allows us to prepare the full range of SKA science cases studying galaxy formation, dynamics, accretion, and depletion processes. This surface brightness sensitivity will be sufficient to trace the faint \ion{H}{i} residing in the outskirts of galaxies and/or intra-group gas.


\subsection{A comparison with other surveys}\label{chap:HI.comp}

\begin{figure}
\centering
\includegraphics[width=7cm]{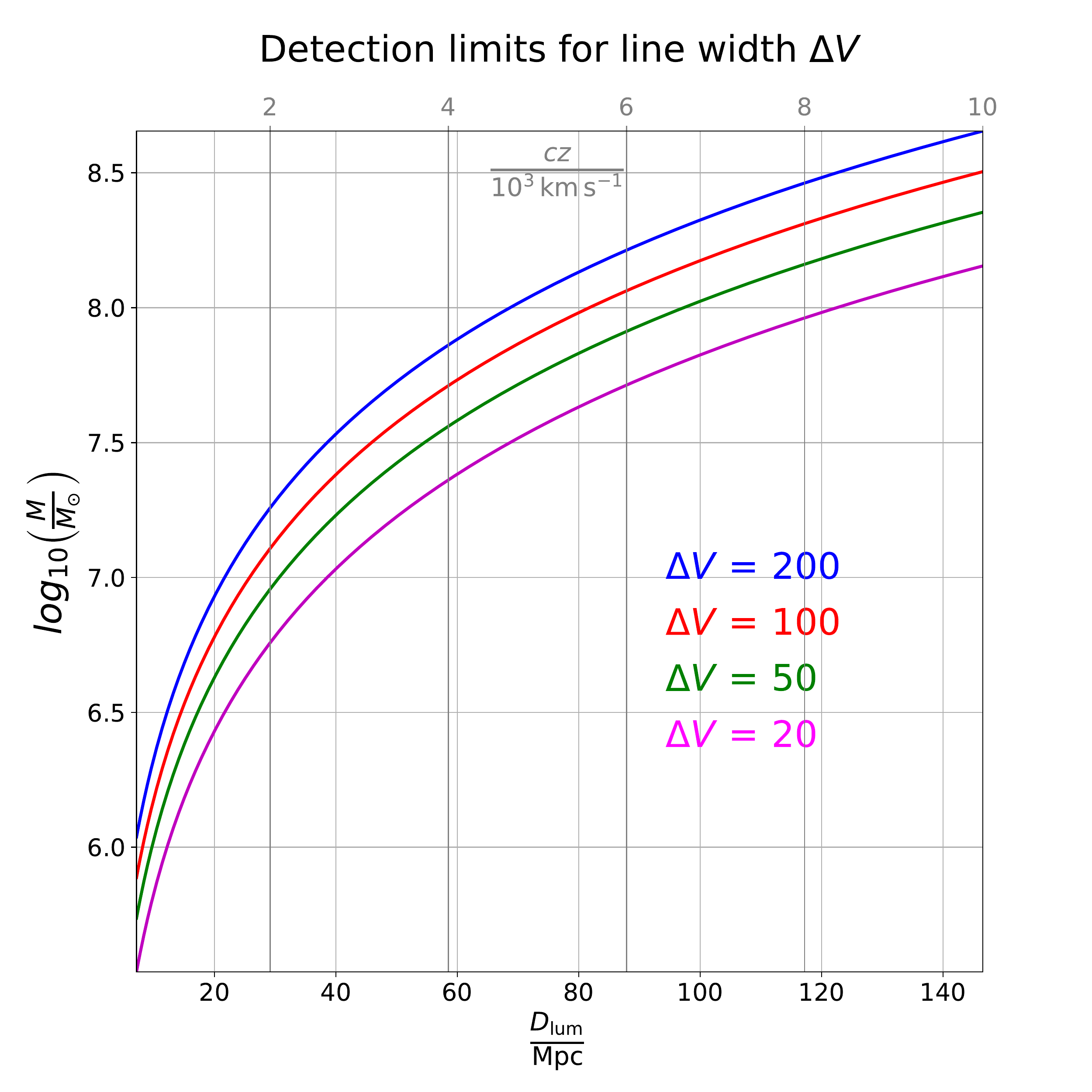}\ \ \ \ 
\includegraphics[width=7cm]{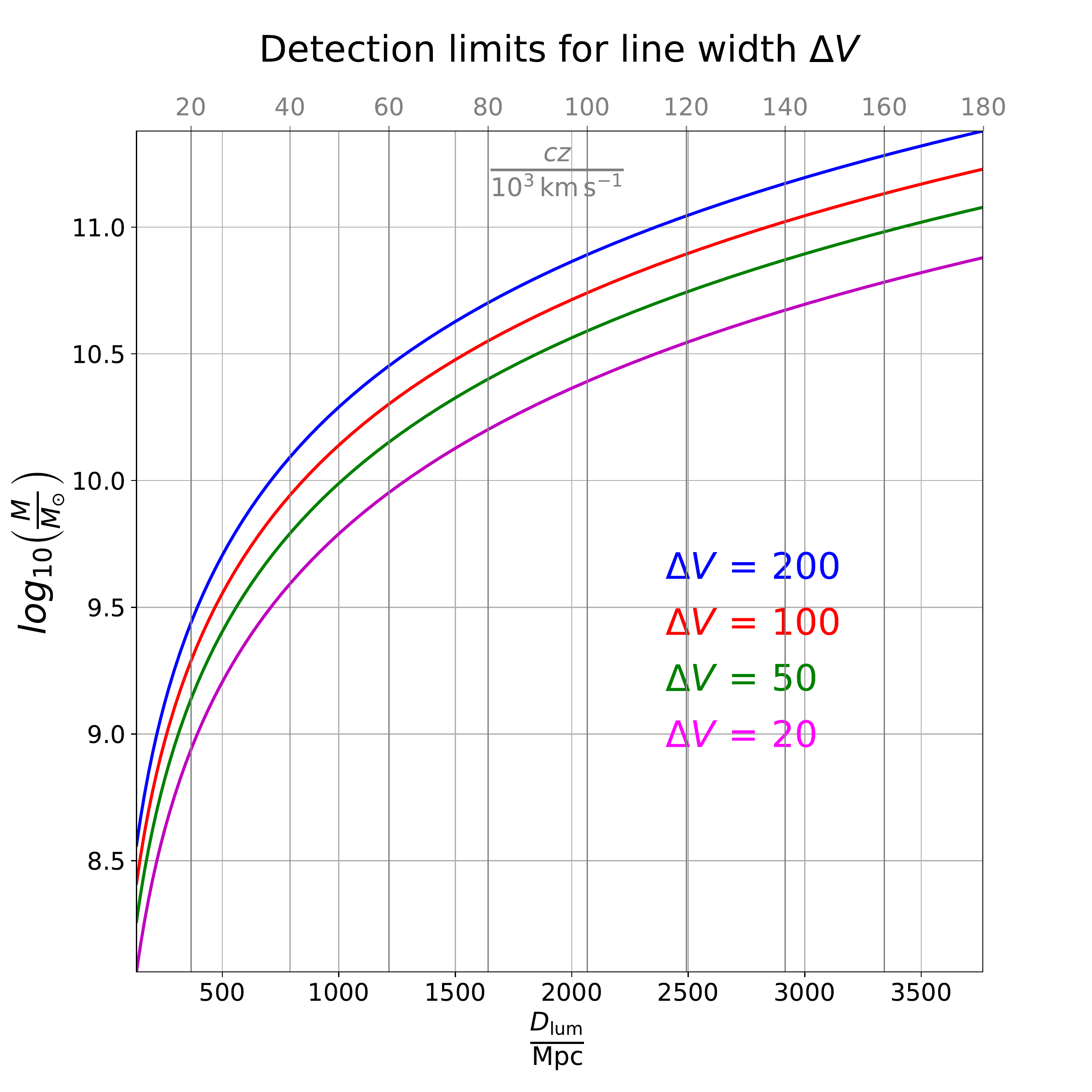}
\caption{\ion{H}{i} point source distance detection limits for MeerKLASS. The panels refer to different redshift ranges. Blue: line width of $200\,\rm km \, s^{-1}$. Red: line width of $100\,\rm km \, s^{-1}$. Green: line width of $50\,\rm km \, s^{-1}$. Magenta: line width of $20\,\rm km \, s^{-1}$.} 
\label{fig:distancedetect}
\end{figure}

With its large survey area and comparably high sensitivity, MeerKLASS will directly detect \ion{H}{i} in a few hundred thousand galaxies. Its sensitivity is a factor of 3-4 better than the current gold standard, ALFALFA \citep{Giovanelli05},
and will hence serve to significantly improve the statistics of integral \ion{H}{i} properties of galaxies. 
Figure \ref{fig:distancedetect} shows the $5-\sigma_{\rm rms}$ point source mass detection limits of MeerKLASS. MeerKLASS will be able to detect the most massive galaxies up to the redshift limits of the MeerKAT L-band receiver (870 MHz, notice that also massive galaxies can have a narrower line width depending on inclination), while objects with low \ion{H}{i} masses, between $10^6\,M_\odot$ and $10^7\,M_\odot$, stay detectable within a reasonable survey volume. A galaxy like Leo T with a dark
matter mass that is just enough to retain baryons, $M(\ion{H}{i})∼2.8\cdot 10^5\,M_\odot$ and radius $∼240$ pc \citep{Ryan-Weber08}, would be detectable at a $5-\sigma_{\rm rms}$ level and resolved up to distances slightly beyond 5 Mpc.
With this, MeerKLASS will be capable of significantly improving statistics of \ion{H}{i} mass, line-width and -shape.

\begin{figure} 
\centering
\includegraphics[width=0.49\textwidth, height=0.35\textwidth]{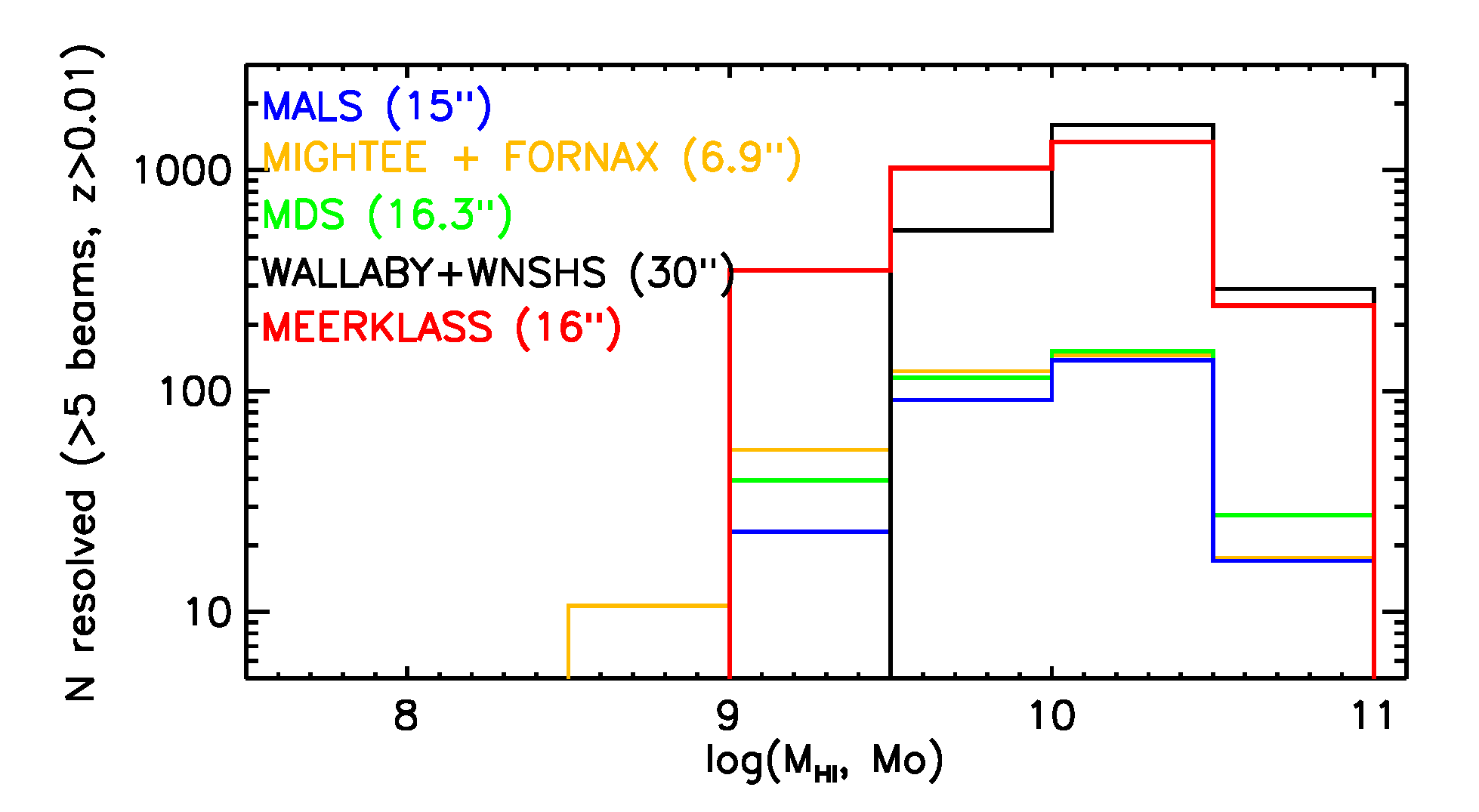} 
\includegraphics[width=0.49\textwidth, height=0.35\textwidth]{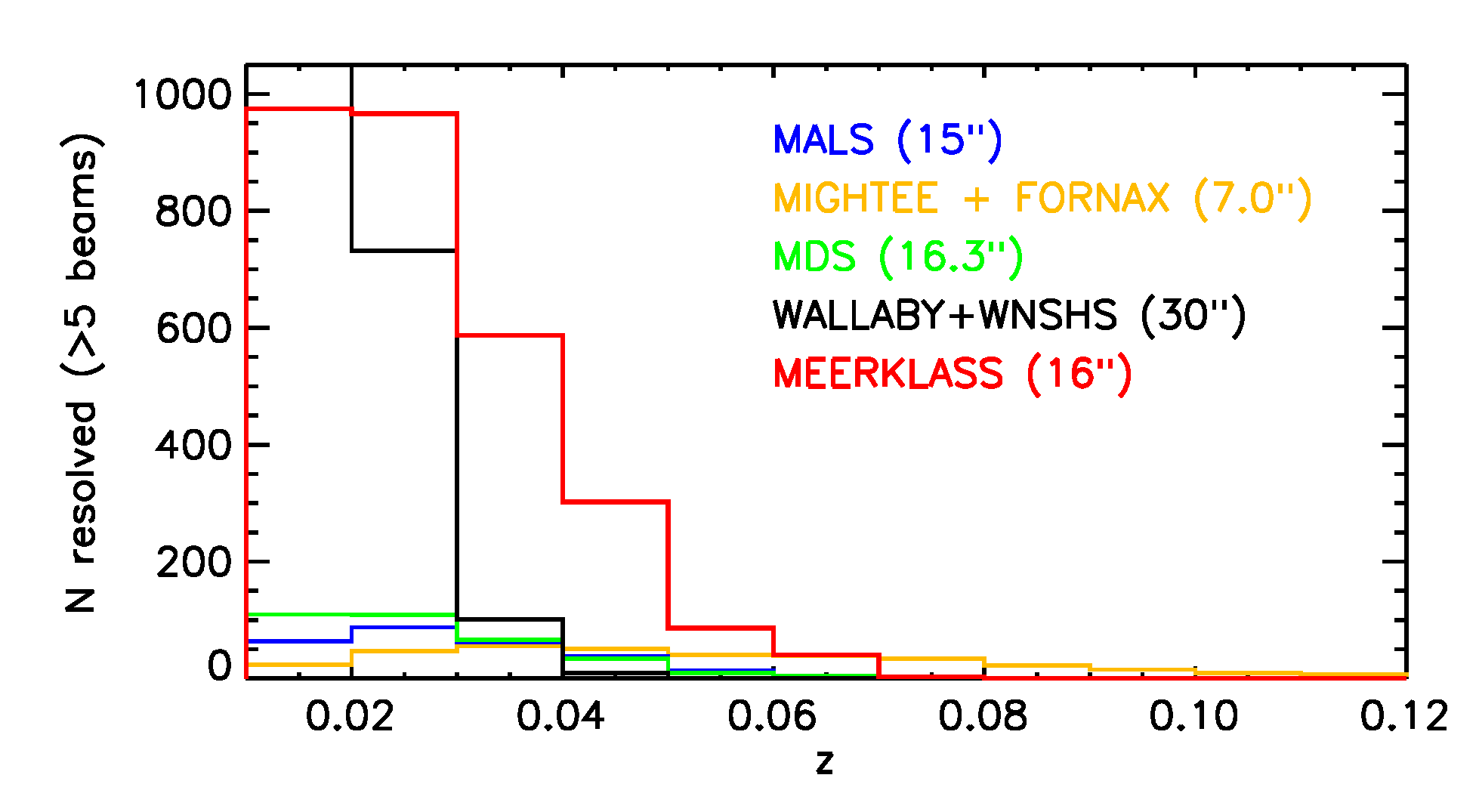}
\caption{Predicted distributions of resolved galaxy masses and redshifts for various surveys beyond $z = 0.01$. {\em Left:} Separated by \ion{H}{i} mass. {\em Right:} Separated by redshift. Blue: MALS, yellow: MIGHTEE and FORNAX combined, green: medium-deep APERTIF, black: WALLABY and APERTIF shallow combined, red: MeerKLASS. Numbers in parentheses show angular resolution.} 
\label{fig:resolved}
\end{figure}

A further indication of the potential impact of MeerKLASS as an \ion{H}{i} survey is evident from Fig.~\ref{fig:resolved}, where its detection statistics for {\it resolved} galaxies at $z > 0.01$ is calculated and compared to various future \ion{H}{i} surveys: we assume a very conservative rms noise of 0.5\,mJy/beam 
over a velocity interval of 20\,km/s 
at $z=0$ (our assumed rms of $5\rm \, \mu Jy$ over the whole L-band, assuming 25\% loss due to RFI, results in a line sensitivity of  $0.4\, \rm mJy\,{\rm beam}^{-1}$). 
This sensitivity corresponds to a 3-$\sigma_{\rm rms}$ detection limit of a column-density of $1\,M_\odot\,\rm pc^{-2}$ (at $z=0$) with a Gaussian beam size of $16^{\prime\prime}$ (HPBW). Such column density limit should be suitable to trace resolved \ion{H}{i} out into the \ion{H}{i}-dominated regimes in galaxies. 

To introduce a metric of the survey impact when targeting resolved objects, we assume a galaxy to be resolved if there are at least 5 synthesized beams across its major axis extent, the latter defined as the location at which the surface density equals $1\,M_\odot\,\rm pc^{-2}$. We predict the \ion{H}{i} extent of each detection using the tight correlation between this quantity and the \ion{H}{i} mass \citep{Broeils97,Wang14,Wang16}, 
the latter estimated from the \ion{H}{i} mass function of \citet{Martin10}. It is clear that MeerKLASS will resolve significantly more galaxies out to larger distances than currently planned \ion{H}{i} surveys, providing a unique sample from which to measure the statistics of disk galaxy structure in the local Universe. 
Even in comparison with a (currently hypothetical) all-sky shallow survey as proposed for ASKAP and APERTIF ("WALLABY+WNSHS"), MeerKLASS will detect a larger amount of galaxies at $z > 0.01$, in particular cutting into the lower mass regime of slightly above $10^9\,M_\odot$, thanks to its superior resolution and hence the ability to also resolve more distant \ion{H}{i} disks at reasonable sensitivity. Adding the very local universe ($0<z<0.01$), we estimate that an all-sky survey with the specs of WALLABY would roughly resolve twice as many galaxies in a mass range below $\log \frac{M_{\rm \ion HI}}{M_\odot}$, but without the ability to trace faint gas.
The Medium-Deep Survey (MDS) with APERTIF is slightly more sensitive compared to MeerKLASS. At $30^{\prime\prime}$ resolution (HPBW), MeerKLASS will reach a $4-\sigma_{\rm rms}$ sensitivity of $4\cdot 10^{19}\,{\rm atoms}\,{\rm cm}^{-2}$ over $20\,\rm km\,s^{-1}$ (here we include a factor of 1.5 in the rms noise estimate to account for losses due to robust weighting), while the MDS will reach $3\cdot 10^{19}\,{\rm atoms}\,{\rm cm}^{-2}$. However, the MDS will not be able to produce comparable statistics; due to the extent of WSRT, its resolving power is limited (to approximately $16^{\prime\prime}$) and cannot be improved by higher sensitivity. Hence, dominated by the ratio of survey areas, MeerKLASS will be able to detect $\approx 9$ times more resolved galaxies than the MDS. 


It should be pointed out that MeerKLASS will be the only \ion{H}{i} survey to both resolve large numbers of galaxies in the mass range between $10^9\,M_\odot$ and $10^{10}\,M_\odot$, on the low-mass side of the knee of the mass function (i.e. rather typical \ion{H}{i} masses of galaxies) and retain the ability to image gas at a low column density level to provide statistics in that mass range. This quality depends heavily on both the rms noise resulting from the integration time, as this guarantees the resolving power (of $16^{\prime\prime}$ HPBW), as well as the survey area, which is 8-10 times larger than foreseen for the medium-deep survey with APERTIF.

\subsection{Neutral hydrogen in galaxies: science}
According to the metrics presented above, MeerKLASS will be superior to any other \ion{H}{i} emission line progenitor survey for the SKA, enabling us to address several major research areas. For example,
the \ion{H}{i} mass function and velocity function are two of the most important measures for gas-rich galaxy populations
\citep{Zwaan03} and can be used to derive the cosmic
\ion{H}{i} mass density; this is an essential input into our understanding of the 
evolution of the cosmic SFR density and the processes governing the
distribution and evolution of cool gas \citep[e.g.][]{Obreschkow09}. With the large volume covered, and its high sensitivity,
MeerKLASS will enable the study of the \ion{H}{i} mass-function as a function of
the environment; the resolution of the MeerKAT means MeerKLASS will be far less hampered by confusion than single-dish surveys or surveys limited by the array extent. MeerKLASS will therefore be highly complementary to, in particular, WALLABY, which will deliver much higher detection statistics, but at an angular resolution of $30^{\prime\prime}$.
In addition to a direct comparison of the \ion{H}{i} mass
function with theoretical predictions \citep[e.g.][]{Power10}, in
combination with complementary surveys at other wavelengths, the
baryonic and total mass function for galaxies can be derived to
further test models of galaxy formation \citep[e.g.][]{Zwaan10}.

MeerKLASS will provide
high-sensitivity data for a large number of well-resolved galaxies (with \ion{H}{i}
diameters in excess of 5 beams), which, in combination with surveys
at other wavelengths (NIR, FIR, H$\alpha$, UV), will enable statistical
studies of the correlation of star formation with the internal \ion{H}{i}
distributions and dynamical properties of galaxies \citep{Bigiel08,Leroy08,Wang14,Wang16}, and challenge the standard cosmology by
testing predictions about galaxy kinematics and dynamics \citep{Blok15}. In nearby groups, galaxies with \ion{H}{i} masses as low as a few $10^5\,M_\odot$ can be
detected.

Due to the comparably high
resolution, MeerKLASS will be able to resolve environmental effects on
galaxies, such as gas stripping \citep[e.g.][]{Oosterloo05}, gas
accretion and galaxy interaction \citep[see e.g.][]{Sancisi08}, thus
probing the mechanisms that regulate star formation. Gas masses of galaxies as a function of galaxy density provide an important insight into the
environmental influences on galaxy evolution. MeerKLASS will be able to reach a sensitivity regime that makes it possible to  characterise outer \ion{H}{i} disks, and the faint surroundings of galaxies on a statistical basis.

Resolved kinematics and morphologies of a large number of galaxies allows us to study gas accretion processes onto galaxies \citep[see e.g.][]{Sancisi08}, manifesting in gaseous halos, warps and lopsidedness, with statistically significant findings. It will also be possible to study the tightly correlated spin characteristics of galaxies in relation to their environment \citep{Obreschkow15}. An additional implication is that it will be possible to search for spin alignment in the outer disks; in an inside-out scenario of galaxy formation, this gas would be accreted most recently in a galaxy's evolution and is thus more likely to be aligned compared to the inner disk \citep{Popping15}. For WALLABY, to reach this goal is already problematic as its column density sensitivity is too low, and for the APERTIF MDS the number statistics are significantly lower. MeerKLASS will therefore be a highly complementary survey to WALLABY, which will provide higher statistics for the inner, high column-density regime.


\subsection{Gas in galaxies: legacy value}
In summary, MeerKLASS will enable us to: (1) establish a \ion{H}{i} mass function, and measure spectral characteristics of unresolved sources with an unprecedented accuracy, even in dense environments, (2) measure radial gas profiles of  
galaxies to study galaxy evolution; 
(3) quantify the matter  
distribution in galaxies; (4) study the intrinsic and mutual spin distribution in galaxies and its   
dependence on environment, and 
(5) study the low  
column density environment of galaxies to quantify AGN feedback and accretion processes on a  
statistical basis. 
Again, MeerKLASS will enable us to observe the \ion{H}{i}-dominated regime beyond the stellar disks, which is highly relevant in the context of galaxy evolution \citep{Wang14, Wang16}, as well as for dynamical studies \citep{Obreschkow15}.

MeerKLASS will produce standard data products to address the above questions, namely: (1) basic emission line  parameters; (2) \ion{H}{i} emission line maps and data cubes of resolved objects; (3) parametrisations of extended morphology and kinematics, spin characteristics, rotation curves, and surface density profiles of resolved galaxies, and (4) velocity and mass functions.

It is currently unclear how the shallow basis of the wedding cake for \ion{H}{i} surveys will be established. Within its anticipated life time, APERTIF will not be used for a shallow survey that covers a quarter of the sky. A real all-sky survey, as assumed in our calculations, will hence not exist on a short time scale and the numbers of resolved galaxies observed by WALLABY and the APERTIF shallow survey will hence be smaller than calculated above. Given the performance of MeerKAT, MeerKLASS could therefore even significantly complement WALLABY and the APERTIF shallow survey by the sheer number of resolved detections to provide high-number statistics for \ion{H}{i} astronomy at low redshifts, in addition to its quality to trace \ion{H}{i} at a far lower column density.

\section{The polarized sky -- cosmic magnetic fields}

\begin{figure}
	\centering
	\includegraphics[width=10cm, height=7cm]{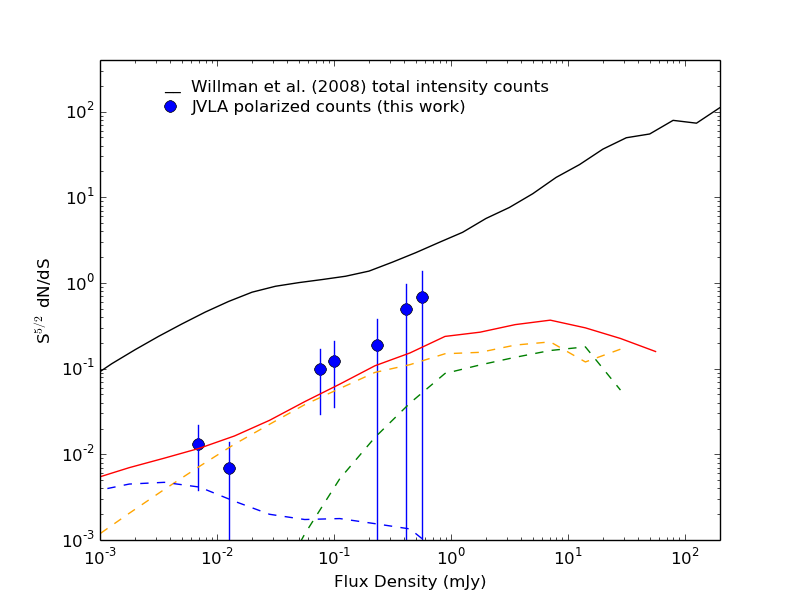}
	\caption{Polarized intensity source counts over  0.1\,deg$^2$  at 5\,GHz \citep{2014ASInC..13...99T}. Solid black line:  
		SKADS simulation of total intensity counts. 
		Red curve: predicted total polarized flux density counts. Dashed curves are predicted counts  for FR II (green), FRI (orange) and normal galaxies (blue).}\label{fig:polcount}
\end{figure}

Current studies of the sub-mJy polarized sky show that it presents questions which cannot be answered by the strong source population alone \citep{2009ASPC..407...12T}.  The mean fractional polarization of radio sources is anti-correlated with flux density.  The faint end of this flux density range contains relatively more objects below the FRI/FRII luminosity boundary than the bright end. This raises a number of questions about the nature and evolution of the faint polarized sources. How is the higher degree of polarization of faint sources related to source structure, radio luminosity, redshift, or environment? Does the trend of increasing polarization continue to lower flux densities? When does it stop? Does the polarization of all AGN increase with flux density, or can we identify a sub-class of AGN that is responsible for this trend?

The fractional polarization and intrinsic polarization angle of a source measure the order and direction of its magnetic field. It is expected that these quantities will be different for distinct classes of object due to the different origin of the emission. In AGN they will mainly be related to the ordered magnetic field in the jets and lobes; for SF galaxies, which are mostly spirals, they are likely to reflect the degree of ordering in the intrinsic disk field. MeerKLASS will have the sensitivity to provide direct detection of the presence and properties of ordered magnetic fields in a vast number of galaxy disks out to intermediate redshifts. At even higher redshifts we will be able to measure statistical polarized source properties to nanoJy levels by stacking the polarized emission at positions corresponding to total intensity detections, and thus trace the emergence of ordered fields in galaxies over cosmic time. With multi-wavelength information we will correlate the radio polarization properties with optical emission line diagnostics, galaxy type and star-formation rates, revealing the relation of the evolution of magnetic fields and the global star formation properties of galaxies.

MeerKLASS in the L-band will directly 
detect thousands of polarized sources spanning the transition where SF and 
normal galaxies become a major component of the polarized population.
MeerKLASS will uniquely provide broad-band spectro-polarimetry for sources at $\mu$y levels over thousands of square degrees.
The large bandwidth will provide high precision Rotation Measure (RM) synthesis making it possible to statistically examine the distribution of angular distribution of RMs to search for the imprint of intergalactic fields on galaxies on cosmological scales. We will also trace the degree of polarization versus frequency due to internal Faraday rotation of the sources. The amount of depolarization by Faraday dispersion sets apart star-forming galaxies from most AGN powered radio sources, potentially allowing us to separate them using the polarization observations; something which can be cross-checked using the multi-wavelength information.  

The bandwidth of the survey will allow Faraday Rotation Measures to be measured with an accuracy better than 2 rad-m$^{-2}$ down to polarized flux densities of 100\,$\mu$Jy. 
From recent JVLA imaging to 1\,$\mu$Jy rms of a 0.1\,deg$^2$ field at 5\,GHz (Fig.~\ref{fig:polcount}), we expect to detect several 10s of sources per square degree down to 30\,$\mu$Jy at about 1.3\,GHz. The 2-point correlation function of RM will be measured on scales of arc minutes and precision of a few  rad-m$^{-2}$, precisely the regime where fluctuations from primordial magnetic fields are expected to create detectable signal \citep{2010ApJ...723..476A}.  This regime of sky density and RM accuracy has not been accessible to date, and will not be surpassed until SKA1.

\section{Summary}

The proposed MeerKAT survey is set to become the leading radio Cosmology survey in preparation towards SKA1-MID. Its design fits well with the MeerKAT set-up both in terms of sensitivity and resolution. The survey will rely both on the dish auto-correlations (single-dish mode) and cross-correlations (interferometer mode). With this, it will provide a 4,000\,deg$^2$ map in HI intensity mapping, continuum ($\sim 5\,\mu$Jy rms) as well as HI line and polarization. This will allow us to achieve the following science cases:
\begin{itemize}
  \item {\bf Cosmology:} {\bf(1)} detect the HI intensity mapping signal both in the auto power spectrum and in cross correlation with other surveys such as DES and CMB experiments (including lensing); {\bf(2)} measure baryon acoustic oscillations and redshift space distortions and corresponding constraints on dark energy and modified gravity theories; {\bf(3)} constraints on primordial non-Gaussianity using the large scale power spectrum and the multi-tracer technique; {\bf(4)} detect about 10 million galaxies in continuum capable of providing dark energy constraints in combination with multi-wavelength data.
\item {\bf Galaxy evolution:} {\bf(1)} evolution of the AGN and SF galaxy populations out to high redshifts; {\bf(2)} measure the environments and clustering of AGN; {\bf(3)} identify high redshift AGN.
\item {\bf Clusters:} {\bf(1)} test  formation and evolution theories for radio halos, relics, and mini-halos via a statistically significant blind cluster sample; {\bf(2)} investigate cluster magnetic fields through Faraday rotation of background sources over a wide range of cluster mass and redshift.
\item {\bf Galaxy HI emission:} {\bf(1)} global galaxy spin distribution and (dark) matter distribution in galaxies; {\bf(2)} gas accretion; {\bf(3)} quenching of star formation.
\item {\bf Polarization:} {\bf(1)} detect thousands of polarized sources spanning the transition where SF and normal galaxies become a major component of the polarized population; {\bf(2)} search for the imprint of intergalactic fields on galaxies on cosmological scales; {\bf (3)} trace the emergence of ordered fields in galaxies over cosmic time; {\bf(4)} detect primordial magnetic fields.
\end{itemize}
Many other science cases will certainly take advantage of such a survey. In particular, if a fast scanning strategy is adopted (on the fly scanning), it will allow us to survey the area quickly, which will be especially suited to look for transients.

Several challenges will have to be overcome, however, if we want to use this signal for cosmological purposes. In particular, cleaning of the huge foreground contamination, removal of any systematic effects and calibration of the system. Foreground cleaning methods for intensity mapping have already been tested with relative success taking advantage of the foreground smoothness across frequency. The planned fast scanning strategy will allow us to deal with the typical $1/f$ noise expected in single-dish surveys by taking advantage of the system stability. Using the same scanning strategy for the interferometer calibration will require sophisticated analysis methods in order to take advantage of the usual self-calibration process. Several techniques are currently being developed and we plan to run several tests with the MeerKAT system in the next few months. Although challenging, this approach has the promise to open a completely new observational window for Cosmology as well as to provide unprecedented high quality data for many other scientific purposes.

\paragraph*{Acknowledgements.}
JF, MH, MJ, RM, OS, MP, MGS, AWi and IW are supported by the South African Square Kilometre Array Project and the National Research Foundation. PB is supported by the NASA Postdoctoral Program at the Jet Propulsion Laboratory, California Institute of Technology, administered by Universities Space Research Association under contract with NASA. DA and PGF are supported by the ERC, UK STFC and the Beecroft Trust. KK is supported by the Claude Leon Foundation. RM is also supported by the UK STFC, Grant ST/N000668/1. AP was supported by a Dennis Sciama Fellowship at the University of Portsmouth. AP acknowledges the University of Western Cape for hospitality.  AWe is supported by the South African Research Chairs Initiative and the National Research Foundation. JA gratefully acknowledges support from the Science and Technology Foundation (FCT, Portugal) through the research grant PTDC/FIS-AST/2194/2012 and PEst-OE/FIS/UI2751/2014.


\bibliographystyle{apj}

\bibliography{meerklass}

\end{document}